\documentclass[a4paper,12pt,final]{article}
\usepackage[flushmargin]{footmisc}
\usepackage{amsmath}
\usepackage{amsfonts}
\usepackage{rotating,graphics,eurosym}
\usepackage{longtable,lscape}
\usepackage[english]{babel}
\usepackage{pgf,pgfarrows,pgfnodes,pgfautomata,pgfheaps,pgfshade}
\usepackage{float}
\usepackage[comma]{natbib}
\usepackage{setspace}
\usepackage{multirow}
\usepackage[dvips]{epsfig}
\usepackage[dvips]{graphicx}
\usepackage{amssymb}
\usepackage{threeparttable}
\usepackage{subfigure}
\usepackage{tabularx}
\usepackage[T1]{fontenc}
\DeclareGraphicsExtensions{.eps .jpg} 

\def\inf{\operatornamewithlimits{inf\vphantom{p}}}
\def\sup{\operatornamewithlimits{sup\vphantom{p}}}

\def\argmin{\operatornamewithlimits{argmin\vphantom{p}}}

\begin{document}

\pagenumbering{arabic} \setlength{\unitlength}{1cm}
\pagestyle{empty} \doublespacing

\title{Heterogeneous earnings effects of the Job Corps by gender:  A translated quantile approach}

\author{Anthony Strittmatter\footnote{
Previous versions of this manuscript were circulated under different titles. The manuscript was presented at the SSES 2012 in Zurich, at the NOeG 2012 in Vienna, at the EALE 2013 in Torino, at the ESEM 2013 in Goteborg, at the IAAE 2014 in London, at the Humboldt University Berlin 2015, and at the ZEW 2017. I thank the participants for their useful comments. In particular, I am grateful for helpful suggestions made by Simone Balestra, Xuan Chen, Annabelle D\"{o}rr, Alfonso Flores-Lagunes, Bernd Fitzenberger, Michael Lechner, Jan Nimczik, Stefan Sperlich, and Andreas Steinmayr. 
I thank Ozkan Eren and Serkan Ozbeklik, who kindly shared their data and programmes with me. The usual disclaimer applies. E-mail:
anthony.strittmatter@unisg.ch.}\\ {\small University of
St. Gallen}}

\date{ 
\today }

\maketitle

\begin{abstract}
Several studies of the Job Corps tend to find more positive earnings effects for males than for females. This effect heterogeneity favouring males contrasts with the results of the majority of other training programmes' evaluations. Applying the translated quantile approach of Bitler, Hoynes, and Domina (2014), I investigate a potential mechanism behind the surprising findings for the Job Corps. My results provide suggestive evidence that the effect of heterogeneity by gender operates through existing gender earnings inequality rather than Job Corps trainability differences. 
\end{abstract}

\noindent \small
JEL-Classification: J38, C21 

\normalsize

\maketitle

\thispagestyle{empty} \setcounter{page}{1} \normalsize

\clearpage \pagestyle{plain}

\section{Introduction}

In this study, I investigate the heterogeneous earnings effects of the Job Corps by gender. In particular, I show that gender differences in labour market opportunities contribute to the overall effect heterogeneity. To this end, I apply the translated quantile decomposition of Bitler, Hoynes, and Domina (2014). Furthermore, I extend this approach to incorporate average translated effects.

The Job Corps is the largest U.S. labour market programme targeting disadvantaged youth. It provides academic, vocational, and social training, as well as health care counselling and job search assistance, for an average duration of eight to nine months. \cite{bl13a}, \cite{chen15b}, and \cite{lee09} show that the Job Corps has positive wage effects for both genders. This effect is expected to increase earnings through higher hourly wages and greater labour supply.\footnote{\cite{chen15} report positive employment effects of the Job Corps four years after assignment.} The female labour supply is expected to be more elastic to wage changes than is the male labour supply \citep[see, e.g., the discussion in][]{bar13,ki86}. However, several studies observe a tendency that the Job Corps increases the gender earnings inequality. In a large-scale experimental evaluation, \cite{schoch08} show evidence of higher average earnings gain from the Job Corps for males than for females. Moreover, \cite{eren14} report more positive quantile earnings effects for males than for females. These findings contrast with the empirical evaluations of many other active labour market programmes (ALMPs). Meta studies performed by \cite{berg08} and \cite{card15} report evidence of higher ALMP returns for females than for males. In this study, I highlight one potential mechanism behind the unexpected findings for the Job Corps.

\cite{fru12} investigate the effect heterogeneity for various socio-economic groups using a principal stratification approach with strong functional form assumptions. They observe that individuals with worse initial conditions (in terms of education, labour market experience, race, and gender) profit less from the Job Corps than do those with good labour market opportunities.\footnote{This is another peculiarity of the Job Corps because for many other ALMPs, the returns do not increase with labour market opportunities \citep[see, e.g.,][]{card15,ri11}.} Expected earnings can be viewed as an aggregate measure of labour market opportunities. Quantile regressions can identify (distributional) heterogeneity by expected earnings. \cite{eren14} report larger earnings differences for the Job Corps at higher quantiles than at lower quantiles. Such a pattern can be expected if the Job Corps is more effective for groups with better labour market opportunities, that is, those with high earnings even without training. Higher opportunity costs for participants with better earnings opportunities can explain this positive relationship between earnings opportunities and Job Corps returns. It is well documented that a larger share of females is located in the lower part than in the upper part of the earnings distribution. Thus, unexpected effect heterogeneity can arise in favour of males from structural gender earnings inequality rather than from gender differences in Job Corps trainability. The latter could arise from programme content that is, for example, more aligned to the needs of one gender. In this study, I decompose those potential channels using the translated quantile approach of \cite{bit13}.

For an example of the relevance of the translated quantile approach, assume the need to determine assignment criteria for offering participation in the Job Corps that do not increase gender inequality.
If structural gender earnings inequality leads to heterogeneity, a non-discriminatory assignment rule offers participation to males and females with balanced labour market opportunities. If Job Corps trainability differs by gender, then other dimensions of the Job Corps programmes need to be better aligned with the specific training demands of males and females. For example, vocational training offered during the programme could meet the needs of males better than the needs of females.

My findings, consistent with the existing literature, provide suggestive evidence for larger quantile treatment effects for males than for females. Structural gender earnings inequality possibly accounts for 82\% of the average effect heterogeneity by gender. Trainability differences by gender appear to play only a minor role. Even though the effects are statistically insignificant, these findings suggest that the offer to participate in the Job Corps does not disfavour females with earnings opportunities that are equal to those of males. However, it seems that the Job Corps amplifies the existing structural earnings inequality in favour of males. Randomly offering participation in the Job Corps to eligible individuals increases the average earnings opportunities by 8\% after four years but could possibly also increase the average gender earnings gap by 8\% within the eligible group. An allocation scheme that balances the earnings structures of the selected programme participants by gender could yield similar average returns but tends to have a lower impact on gender earnings inequality. 

Furthermore, I report effect heterogeneity by gender and parenthood. While mothers tend to profit more from Job Corps than do fathers, the results suggest the opposite for women and men without children. I document that differences in the earnings structure by gender and parenthood can account for a large proportion of this heterogeneity (on average, 71\%). This suggests the within-group earnings structure is an important factor for between-group effect heterogeneity. Accordingly, the within-group earnings structures should receive more attention when rules used to award Job Corps offers are designed.

The remainder of the paper is structured as follows. Section 2 reviews the institutional background of the Job Corps and the related literature. Section 3 describes the experimental data. Section 4 explains the quantile transformation approach. Section 5 presents the main empirical results. Section 6 describes effect heterogeneity by gender and parenthood, and the final section concludes the paper. Online Appendix \ref{app_new} contains supplementary descriptive statistics. Online Appendix \ref{appA} provides the details on the estimation approach. Online Appendix \ref{appB} presents robustness checks.
Online Appendix \ref{appE} documents distributional effect heterogeneity by gender and parenthood.

\section{Job Corps}

The Job Corps was established in the United States in 1964 under the Economic Opportunity Act. The aim of the Job Corps is to provide counselling and to develop individualized programmes that enhance participants' job opportunities. In particular, completing a vocational degree and earning General Educational Development (GED) credentials are two major goals of the Job Corps. Job Corps programmes follow a holistic approach. They can include physical and mental healthcare counselling, various vocational and social education schemes, and job search assistance. Courses have theoretical and practical content, and training can be partly on the job. Participants are trained in one of approximately 125 Job Corps centres and typically live on Job Corps campuses during the programme. Candidates must pass a screening and recruitment process administered by certain admission agencies. Eligible participants are disadvantaged youth (aged 16--24 years) in need of additional education and training. Candidates must have low incomes to meet the eligibility criteria. Approximately 60,000 eligible youth enrol in the Job Corps each year. In 2008, total programme expenditures exceeded 1.58 billion U.S. dollars 
\citep{jc08}.\footnote{Of the total expenditures, training costs account for 40\%, support services (e.g., meals, lodging, and medical care) account for 33\%, and administration and infrastructure investment costs account for 26\%.} In 2008, expenditures per participant were approximately 26,000 U.S. dollars, or approximately ten times the cost of a typical programme under the Workforce Investment Act \citep{eren14}.

Numerous studies investigate the effectiveness of Job Corps programs in improving participants' post-programme earnings. Early studies using data from the 1970s observe that males benefited more from the Job Corps than did females. \cite{ma82} observe that participation in the Job Corps increases annual earnings by 2,000 U.S. dollars for males and 1,000 U.S. dollars for females. \cite{ga80} report negative earnings impacts for black female participants and positive effects for black male participants. Between 1994 and 1996, Mathematica Policy Research performed a randomized National Job Corps Study (NJCS). That study had an experimental design and hence became the benchmark for evaluating the Job Corps. Using survey data collected from NJCS participants in 1998, \cite{schoch08} observe a yearly earnings gain of 1,530 U.S. dollars for males, but only 1,134 U.S. dollars for females. These earnings differences are not statistically significant. However, \cite{schoch08} report a significantly higher yearly earnings gain of 540 U.S. dollars for male than for female participants when using the administrative data of NJCS participants collected from the 1998 Summary Earnings Records (SER). \cite{flo12} explore the marginal returns of additional exposure to academic and vocational training and instruction by gender. The average marginal returns of additional exposure are twice as large for males than for females.

Furthermore, \cite{schoch08} report the effects of receiving an offer to participate in the Job Corps, which could include actual participation in the programme. Using the survey data collected from NJCS participants four years after randomization, the authors observe an average yearly earnings gain of 828 U.S. dollars, equivalent to an average weekly earnings gain of 16 U.S. dollars. Using the same data, \cite{eren14} report weekly earnings gains of 18 U.S. dollars for male and 11 U.S. dollars for female recipients of an offer to join the Job Corps. Quantile differences suggest almost zero effects on the earnings distribution of males below the 4th decile. Above the 4th decile, the quantile earnings differences increase to approximately 30 U.S. dollars and remain at this positive level. For females, the quantile earnings differences reach approximately 20 U.S. dollars at the 15th percentile. Above this percentile, the positive quantile earnings differences remain almost stable. Thus, offers to join the Job Corps tend to increase gender earnings inequality. 

\section{Experimental data}

As previously mentioned, Mathematica Policy Research performed a randomized experiment with the Job Corps. Of 80,833 eligible individuals who first applied for participation in the programme between November 1994 and February 1996, 15,386 were randomly selected for the experiment's research sample. From this group, 9,409 were randomly selected to receive an offer to participate in the Job Corps. Of these, 73\% joined the Job Corps programme and actually started to participate after an average duration of 1.4 months. The remaining 5,977 individuals were assigned to the control group. Control group members were not eligible for the Job Corps programs for three years following randomization (approximately 1\% participated anyway). Nevertheless, most control group members received some sort of alternative training in the form of GED preparation (27\%), high school (23\%), or vocational, technical, or trade school (21\%) (Burghardt et al., 2001)\nocite{sch01b}. The intensity of training was higher among treatment group members than among those in the control group. During the four years following the randomization, 47\% of the treatment group members and 27\% of those in the control group obtained GED credentials. During the same period, 38\% of the treatment group members and 15\% of those in the control group obtained a vocational degree. Approximately 5--8\% of eligible individuals in both groups received a high school diploma (Burghardt et al., 2001)\nocite{sch01b}. To improve the accuracy of the research sample, some subgroups were oversampled \citep[see description in][]{schoch01}. To account for this, I use sampling weights throughout the analyses.\footnote{\cite{eren14} condition on key features of the sample design in a regression framework instead of using sampling weights. Unreported robustness checks find that the results are not sensitive to various ways of constructing sampling weights.}

Interviews were conducted with members of the research sample 12, 30, and 48 months after randomization. This study considers publicly released interviews of the sample of individuals conducted 48 months after randomization, eliminating observations with missing information (3,951 observations are dropped).\footnote{3,650 observations are dropped because earnings in year 4 are unobserved, and 301 observations are dropped because of missing data of covariates.}  The share of attrition is 6.5\% among males and 6.1\% among females. \cite{lee09} proposes an indirect test for potential attrition bias in the NJCS data. His analysis reveals no strong evidence of attrition bias \citep[see Remark 2 in][]{lee09}. Throughout the study, I use the non-response adjusted sampling weights (\textit{wgt48b}) provided in the NJCS data to account for sample attrition. The final sample includes 10,595 observations, of which 6,372 are in the treatment sample and 4,223 are in the control sample. 

The outcome of interest is average weekly earnings (in nominal U.S. dollars) in year 4 after random assignment, which is more than two years after 92\% of the participants completed the program.
I investigate the effects of the randomized offer to participate in the Job Corps programme. This intention-to-treat effect (ITT) differs from the effect of actual participation. The ITT effect is relevant to investigating the optimal selection rules for awarding offers to join the Job Corps programme. The award of such offers is under the control of the decision makers; however, enrolment and actual participation are not under their control. An alternative approach is to investigate the local average treatment effects of actual participation. This approach identifies the effects for those individuals who only enrol in Job Corps if they receive the offer to participate instead of considering all eligible individuals (the latter group is called compliers). \cite{chen15} address this limitation with an interesting bound analysis designed to identify the effects for individuals rejecting the offer to participate. I focus only on the ITT effects in this study.

\begin{table}
\caption{Share of compliers.} \label{comp} \footnotesize
\begin{center}
\begin{tabular}{lcccccc}
\hline
\hline
           &      &       &      Share of & Share of &      Share of &      Share of \\

           &     Share of &      Share of &     female &       male &     female &       male \\

           &  female &  male &  compliers &  compliers &  compliers &  compliers \\

           &      compliers      &    compliers        &       with &       with &    without &    without \\

           &            &            &   children &   children &   children &   children \\
	\cline{2-7}				& (1) & (2) & (3) & (4) & ( 5) & (6) \\
\hline
Full sample &       69\% &       75\% &       64\% &       67\% &       71\% &       76\% \\

$Y(0) < 1^{st}$ quartile   & \multirow{1}{*}{69\%} & \multirow{ 1}{*}{75\%} & \multirow{ 1}{*}{64\%} & \multirow{ 1}{*}{62\%} & \multirow{ 1}{*}{71\%} & \multirow{ 1}{*}{76\%} \\

$1^{st}$ quartile $\leq Y(0) < 2^{nd}$ quartile  & \multirow{ 1}{*}{69\%} & \multirow{ 1}{*}{76\%} & \multirow{ 1}{*}{64\%} & \multirow{ 1}{*}{67\%} & \multirow{ 1}{*}{71\%} & \multirow{ 1}{*}{77\%} \\

$2^{nd}$ quartile $\leq Y(0) < 3^{rd}$ quartile & \multirow{ 1}{*}{69\%} & \multirow{ 1}{*}{74\%} & \multirow{1}{*}{64\%} & \multirow{ 1}{*}{66\%} & \multirow{ 1}{*}{71\%} & \multirow{ 1}{*}{76\%} \\

$ Y(0) \geq 3^{rd}$ quartile   & \multirow{ 1}{*}{67\%} & \multirow{ 1}{*}{73\%} & \multirow{ 1}{*}{62\%} & \multirow{ 1}{*}{65\%} & \multirow{1}{*}{69\%} & \multirow{ 1}{*}{74\%} \\
\hline
\hline
\end{tabular}  
\end{center}
\parbox{17cm}{\footnotesize Note: See \cite{ang13} for a description of the calculation of complier shares. The potential outcome $Y(0)$ denotes earnings under non-treatment. }
\end{table}

Table \ref{comp} documents the share of compliers by gender and by different quartiles of the earnings distribution. Females comply 6 percentage points less frequently with the offer to participate in the Job Corps than do males. The difference is statistically significant at all conventional levels. Accordingly, compliance with the offer to join the Job Corps can explain some amount of effect heterogeneity by gender, but the effects of actual participation are also more beneficial for males than for females \citep[see the results in, e.g.,][]{schoch08,flo12}. This suggests that non-compliance cannot explain why participation in Job Corps tends to increase gender inequality of earnings. The share of compliers does not vary significantly across the earnings distribution. Furthermore, parenthood is a strong predictor of compliance for both males and females.

\begin{table}
\caption{Means and standard deviations conditional on receiving (or not) an offer to join Job Corps (JC).} \label{tab1} \footnotesize
\begin{center}
\begin{tabularx}{17cm}{Xccccc}
\hline
\hline
           & \multicolumn{ 2}{c}{Offer to join JC} & \multicolumn{ 2}{c}{Not invited to JC} & Standardized \\

           &       Mean &  Std. Dev. &       Mean &  Std. Dev. & Difference \\

           &        (1) &        (2) &        (3) &        (4) &        (5) \\
\hline
Earnings per week in Year 4 &    210.854 &    200.359 &    195.846 &    187.475 &      7.740 \\

Ever enrolled in a Job Corps centre &      0.733 &      0.442 &      0.011 &      0.106 &    224.320 \\
\hline
                  \multicolumn{ 6}{c}{{\bf Socioeconomic characteristics}} \\
\hline
    Female &      0.412 &      0.492 &      0.406 &      0.491 &      1.200 \\

Aged 16-17 years &      0.414 &      0.493 &      0.417 &      0.493 &      0.590 \\

Aged 18-19 years &      0.317 &      0.465 &      0.317 &      0.465 &      0.070 \\

Aged 20-24 years &      0.269 &      0.443 &      0.266 &      0.442 &      0.580 \\

     White &      0.277 &      0.447 &      0.266 &      0.442 &      2.470 \\

     Black &      0.479 &      0.500 &      0.478 &      0.500 &      0.120 \\

  Hispanic &      0.171 &      0.377 &      0.180 &      0.384 &      2.420 \\

    Native American &      0.074 &      0.261 &      0.076 &      0.265 &      0.900 \\

Lives with a spouse or a partner &      0.061 &      0.239 &      0.063 &      0.243 &      0.980 \\

Dummy for bad health &      0.127 &      0.333 &      0.137 &      0.344 &      3.090 \\

0-6 month education programme in the preceding year &      0.268 &      0.442 &      0.254 &      0.435 &      2.990 \\

6-12 month education programme in the preceding year &      0.352 &      0.477 &      0.368 &      0.482 &      3.360 \\

High school credential &      0.228 &      0.420 &      0.238 &      0.426 &      2.380 \\

Lives in PMSA &      0.319 &      0.466 &      0.327 &      0.469 &      1.610 \\

Lives in MSA &      0.459 &      0.498 &      0.454 &      0.498 &      1.110 \\
\hline
            \multicolumn{ 6}{c}{{\bf Past employment and earnings history}} \\
\hline
Ever had job for two or more weeks &      0.803 &      0.398 &      0.792 &      0.406 &      2.750 \\

Worked during the year prior to random assignment &      0.653 &      0.476 &      0.645 &      0.478 &      1.750 \\

Had a job at the time of random assignment &      0.211 &      0.408 &      0.206 &      0.404 &      1.300 \\

Employed less than 3 months during the preceding year  &      0.190 &      0.380 &      0.189 &      0.379 &      0.270 \\

Employed 3-9 months during the preceding year  &      0.282 &      0.437 &      0.276 &      0.433 &      1.510 \\

Employed 9-12 months during the preceding year &      0.181 &      0.373 &      0.180 &      0.372 &      0.230 \\

Yearly earnings less than \$1,000 &      0.107 &      0.309 &      0.111 &      0.314 &      1.470 \\

Yearly earnings \$1,000 to \$5,000 &      0.273 &      0.446 &      0.264 &      0.441 &      2.060 \\

Yearly earnings \$5,000 to \$10,000 &      0.138 &      0.345 &      0.133 &      0.339 &      1.530 \\

Yearly earnings above \$10,000  &      0.067 &      0.250 &      0.067 &      0.250 &      0.060 \\
\hline
                            \multicolumn{ 6}{c}{{\bf Past welfare history}} \\
\hline
Family on welfare when growing up &      0.200 &      0.400 &      0.193 &      0.395 &      1.530 \\

Received food stamps during the preceding year &      0.435 &      0.496 &      0.435 &      0.496 &      0.090 \\

Public or rent-subsidized housing &      0.204 &      0.400 &      0.193 &      0.392 &      2.940 \\

Received AFDC in last year &      0.300 &      0.458 &      0.297 &      0.457 &      0.620 \\
\hline
                                 \multicolumn{ 6}{c}{{\bf Drugs and crime}} \\
\hline
Used hard drugs during the preceding year &      0.068 &      0.251 &      0.063 &      0.243 &      1.970 \\

Smoked marijuana during the preceding year &      0.249 &      0.432 &      0.243 &      0.429 &      1.380 \\

``Ever arrested'' dummy &      0.250 &      0.433 &      0.253 &      0.434 &      0.700 \\
\hline
No. of Observations &      \multicolumn{ 2}{c}{6,372}           &      \multicolumn{ 2}{c}{4,223}          &            \\
\hline
\hline

\end{tabularx}  
\end{center}
\parbox{17cm}{\footnotesize Note: Time-varying control variables are measured at the time of random assignment. Weights ($wgt48b$) accounting for the sampling design and non-response of the 48-month interview are used. JC denotes for Job Corps. PMSA stands for Primary Metropolitan Statistical Area. MSA denotes Metropolitan Statistical Area. AFDC stands for Aid to Families with Dependent Children. }
\end{table}

Table \ref{tab1} reports the means and standard deviations of the observed variables. In year four after a randomized assignment, individuals with an offer to join the Job Corps earn on average 15 U.S. dollars per week more than do individuals without an offer. Furthermore, I report the descriptive statistics for socioeconomic characteristics, employment, earnings, and welfare histories, as well as drug use and arrests. For all of these observed control variables, I observe little difference between individuals with and without an offer to join the Job Corps. The standardized differences are always below 3.5.\footnote{The standardized difference of variable $X$ between samples $A$ and $B$ is defined as 
\begin{equation*}
SD = \frac{|\bar{X}_A-\bar{X}_B|}{\sqrt{\frac{1}{2}\left(Var(X_A) + Var(X_B)\right)}}\cdot 100,
\end{equation*}
where $\bar{X}_A$ denotes the mean of sample $A$, and $\bar{X}_B$ denotes the mean of sample $B$. \cite{ro83} consider an absolute standardized difference exceeding 20 to be ``large.''}
This confirms that the offer to join the Job Corps was randomized appropriately. 

Table \ref{tab2} in Online Appendix \ref{app_new} documents the descriptive statistics of observed variables by gender without conditioning on an offer to join the Job Corps. On average, males earn 66 U.S. dollars per week more than do females four years after the randomized offer. Relative to males, females in the sample were less likely to be white, more likely to have high school credentials, less likely have used drugs during the previous year, and arrested less often. Females received food stamps and aid to families with dependent children (AFDC) more often than did males. The families of females in the sample were on welfare during their childhood more often than were the families of males. These differences in socioeconomic characteristics may explain effect heterogeneity by gender but cannot explain why the findings for Job Corps differ from those of the previous literature \citep[e.g.,][]{card15}.\footnote{Furthermore, \cite{huber15} points at causal pitfalls in controlling for post-birth covariates in gender decompositions.}

\section{Empirical approach}

\subsection{Econometric Intuition}

The earnings are a function of a Job Corps offer $D$, gender $G$, and an error term $U$ that incorporates labour market opportunities. The structural earnings function is $\varphi_{1}(G,U)$ if there is an offer to join the Job Corps and $\varphi_{0}(G,U)$ if there is no offer. The error term possibly interacts non-separably with gender. The quantile earnings function is $Q_{\varphi_{d}(G,U)}(\tau )$ for $d \in \{0,1\}$.\footnote{$Q_{A}(\tau)$ represents the $\tau$th quantile of $A$.} For an insight, assume that $\varphi_{d}(\cdot)$ is strictly increasing in its second argument.
Then, the conditional quantile function $Q_{\varphi_{d}(G,U)|G}(\tau|g)$ equals $\varphi_{d}(g,Q_{U|G}(\tau|g))$, where $Q_{U|G}(\tau|g)$ is the quantile of the error term distribution for those with gender $g \in \{m,f\}$.
The conditional quantile treatment effects (CQTEs) for males ($g=m$) and females ($g=f$) are, respectively,
\begin{align*}
 \delta_{CQTE}(\tau,m)= &\varphi_1(m,Q_{U|G}(\tau|m))- \varphi_0(m,Q_{U|G}(\tau|m)) \mbox{ and } \\
 \delta_{CQTE}(\tau,f) = &\varphi_1(f,Q_{U|G}(\tau|f))- \varphi_0(f,Q_{U|G}(\tau|f)).
\end{align*}
Thus, even if the earnings functions are homogeneous by gender (i.e., $ \varphi_{d}(m,U) = \varphi_{d}(f,U)$), the CQTEs can be heterogeneous if $Q_{U|G}(\tau|m) \neq Q_{U|G}(\tau|f)$ (this is noted in Abadie, Angrist, and Imbens, 2002 and Bitler, Hoynes, and Domina, 2014, among others).\nocite{ab02,bit13} The direct effect operates through heterogeneity caused by the first argument of the earnings functions. The structural effect operates through heterogeneity caused by the second argument. The direct and structural channels cannot be distinguished simply by comparing $\delta_{CQTE}(\tau,m)$ and $\delta_{CQTE}(\tau,f)$.

\cite{bit13} use unique reference quantiles for their translated quantile treatment effect (TQTE) approach that I denote by $Q_{U_r}(\tau)$. They use the same reference quantile to transform CQTE of different groups:\footnote{\cite{po16} proposes a generalized quantile regression (GQR) that also uses a transformation of conditional quantiles. The main purpose of GQR is to estimate unconditional quantiles under unrestrictive assumptions.} 
\begin{align*}
 \delta_{TQTE}(\tau,m)= &\varphi_1(m,Q_{U_r}(\tau))- \varphi_0(m,Q_{U_r}(\tau)) \mbox{ and } \\
 \delta_{TQTE}(\tau,f) = &\varphi_1(f,Q_{U_r}(\tau))- \varphi_0(f,Q_{U_r}(\tau)).
\end{align*}
Heterogeneity arises between $\delta_{TQTE}(\tau,m)$ and $\delta_{TQTE}(\tau,f)$ solely from the first argument of the earnings function. Accordingly, this approach enables us to isolate effect heterogeneity by gender, but the size of the effect will depend on the reference quantile. I propose using the reference quantiles from the potential earnings distribution under non-treatment. I show that TQTE equals CQTE when no structural earnings inequality exists. Furthermore, I show that TQTE equals the (unconditional) quantile treatment effect (QTE) when structural inequality is solely responsible for the effect heterogeneity by gender. 

The conditional average treatment effects (CATE) are defined as
\begin{equation*}
 \delta_{CATE}(m)= \int_0^1 \delta_{CQTE}(\tau,m) d\tau \mbox{ and } 
 \delta_{CATE}(f) = \int_0^1 \delta_{CQTE}(\tau,f) d\tau.
\end{equation*}
This definition implies that inequalities in the earnings structure can be carried through to average effects. Therefore, I extend the approach in \cite{bit13} to incorporate average effects.

\subsection{Treatment effect framework}

In the programme evaluation literature, it is common to 
use the potential outcomes instead of the earnings functions. The potential earnings under assignment to and possible participation in the Job Corps are $Y_i(1) = \varphi_1(G_i,U_i)$ (for $i = 1, ..., N$). The potential earnings when individual $i$ does not receive an offer to participate in the Job Corps are $Y_i(0)= \varphi_0(G_i,U_i)$. For each
individual, only the realized earnings,
\begin{equation*}
Y_i = Y_i(1) \cdot D_i + Y_i(0) \cdot (1-D_i),
\end{equation*}
are observed, implying that either $Y_i(1)$ or $Y_i(0)$ is
counterfactual.\footnote{The observational rule holds only under the stable unit treatment value assumption \citep[e.g.,][]{rub05}.} The individual-specific causal effects are
\begin{equation*}
\delta_i=Y_i(1)-Y_i(0) = \varphi_1(G_i,U_i) -\varphi_0(G_i,U_i).
\end{equation*}
In the remainder of the paper, I use the potential outcome notation. We cannot identify individual-specific causal effects using cross-sectional data without making strong assumptions, such as rank preservation. The latter implies that individuals do not change their rank because of their treatment status \citep[see discussion in][]{fir07}.\footnote{\cite{ch05} use the term rank invariance to refer to this assumption. The authors also introduce rank similarity, which is a weaker assumption than rank preservation because it
allows for random deviations in ranks.}

Nevertheless, quantile differences can be identified without assuming rank preservation. The potential quantile $Q_{Y(d)}(\tau)$ is the minimum value of $Y_i(d)$ such that at least the share $\tau$ of the potential earnings distribution lies below this value.\footnote{$F_{Y(d)}(y) = Pr(Y_i(d) \leq y)= Pr(Y_i \leq y|D_i=d)$ is identified from observable data because of the random treatment assignment. It denotes the cumulative distribution function of $Y_i(d)$. The $\tau$th quantile of $Y_i(d)$ is
\begin{equation*}
Q_{Y(d)}(\tau) = \inf \{y: F_{Y(d)}(y) \geq \tau \},
\end{equation*}
for the real $\tau \in (0,1)$.} Quantile treatment effects (QTEs) are defined as
\begin{equation*}
\delta_{QTE}(\tau)= Q_{Y(1)}(\tau) - Q_{Y(0)}(\tau),
\end{equation*}
the difference between the potential quantiles. QTEs identify the horizontal differences in potential earnings distributions at specific ranks. 

Furthermore, the potential conditional quantile $Q_{Y(d)|G}(\tau|g)$ is the minimum value of $Y_i(d)$ such that at least the share $\tau$ of the conditional potential earnings distribution lies below this value.\footnote{$F_{Y(d)|G}(y|g) = Pr(Y_i(d) \leq y|G_i=g)= Pr(Y_i \leq y|D_i=d,G_i=g)$ is identified from observable data because of the random treatment assignment. It denotes the potential earnings distribution of $Y_i(d)$ conditional on gender $g$ for $g \in \{m,f\}$. The $\tau$th quantile of $Y_i(d)$ conditional on gender is
\begin{equation*}
Q_{Y(d)|G}(\tau|g) = \inf \{y: F_{Y(d)|G}(y|g) \geq \tau \}.
\end{equation*}} The CQTEs are defined as
\begin{equation*}
\delta_{CQTE}(\tau,g)= Q_{Y(1)|G}(\tau|g) - Q_{Y(0)|G}(\tau|g),
\end{equation*}
the difference between the conditional potential quantiles.

Finally, the average treatment effect (ATE) is the expected value of $\delta_i$,
\begin{equation*}
\delta_{ATE} = \int_0^1 \delta_{QTE}(\tau) d\tau =E[Y_i(1)-Y_i(0)],
\end{equation*}
and the CATE is
\begin{equation*}
\delta_{CATE}(g) = \int_0^1 \delta_{CQTE}(\tau,g) d\tau =E[Y_i(1)-Y_i(0)|G_i=g].
\end{equation*}
Both average effects can be identified without assuming rank preservation.

\subsection{Quantile transformations \label{sec}}

One non-separable method of incorporating the earnings structure is to fix the value of the non-treated outcome at which CQTEs are measured. This requires a unique reference distribution, the quantiles of which are used as fixed anchor points. The CQTEs of various groups are transformed to these reference quantiles to place each CQTE on the same absolute scale. The required quantile transformations can be conceptualized through a relative rank \citep[see][]{hand94}. Although the choice of reference distribution is obviously crucial, generally, there is no best choice. I propose using the potential earnings distribution under non-treatment as a reference distribution and define the relative rank as
\begin{equation*}
\tau_{g}^{r}  = F_{Y(0)|G}(Q_{Y(0)}(\tau)|g).
\end{equation*}
This relative rank can be interpreted as the proportion of the conditional population with potential earnings below the $\tau^{th}$ potential earnings quantile of the population under non-treatment. Subsequently, I demonstrate why this relative rank allows a meaningful decomposition of CQTE. Nevertheless, many alternative measures are available for the relative rank, and these can also be meaningful. Currently, \cite{bit13} is the only other study that uses the translated quantile approach. They use the observed outcome distribution under non-treatment as reference distribution in their instrumental variable approach.

The translated quantile treatment effect (TQTE) is defined as
\begin{equation} \label{ff}
\delta_{TQTE}(\tau,g) = Q_{Y(1)|G}(\tau_{g}^{r}|g) - Q_{Y(0)|G}(\tau_{g}^{r}|g),
\end{equation}
implying that TQTE measures the horizontal distance
between two conditional potential outcome distributions at $\tau_{g}^{r}$. The only difference between CQTE and TQTE is that
the location has shifted from $\tau$ to $\tau_{g}^{r}$. TQTEs establish a mapping between the conditional and reference distributions.

If $F_{Y(0)|G}(y|g)$ is strictly increasing in the interval $(0,1)$, such that $Q_{Y(0)|G}(F_{Y(0)|G}(y|g)|g) = y$, then TQTE can be rearranged as\footnote{For example, suppose that the median potential earnings under non-treatment are 200 U.S. dollars for females. For further simplification, assume that $F_{Y(0)|G}(200|f) = 0.5$ (i.e., the outcome is continuous). If $F_{Y(0)|G}(200|f)$ is strictly increasing, then $Q_{Y(0)|G}(0.5|f) = 200$ is an equality. Otherwise, $Q_{Y(0)|G}(0.5|f) \leq 200$ is not a strict inequality. Both results hold also when the outcome is not continuous, i.e., if $F_{Y(0)|G}(200|f) \leq 0.5$.} 
\begin{equation} \label{eq1}
\delta_{TQTE}(\tau,g) = Q_{Y(1)|G}(\tau_{g}^{r}|g) - Q_{Y(0)}(\tau).
\end{equation}
This quantity is related to the changes-in-changes estimator proposed by \cite{at06}. The structural conditional quantile treatment effect (SQTE) can be defined as 
\begin{equation*}
\delta_{SQTE}(\tau,g) = \delta_{CQTE}(\tau,g) - \delta_{TQTE}(\tau,g). \\
\end{equation*}
This describes the difference between CQTE and TQTE.

For an explicit example, assume that the median potential outcome under non-treatment is 300 U.S. dollars, i.e., $Q_{Y(0)}(0.5)= 300$. However, females earning 300 U.S. dollars would already be at the 75\% rank in the conditional potential earnings distribution of females under non-treatment, i.e., $F_{Y(0)|G}(300|f) =0.75$. Then, the relative rank for females would be $\tau_{f}^{r}  = 0.75$. The median TQTE would measure the difference between the conditional potential quantiles under treatment and non-treatment of females at rank $\tau_{f}^{r}  = 0.75$. In contrast, the median CQTE would measure the difference between the conditional potential quantiles under treatment and non-treatment of females at rank $\tau  = 0.5$. The median SQTE would measure the difference between the conditional potential quantiles under treatment of females at ranks $\tau  = 0.5$ and $\tau_{f}^{r}  = 0.75$. Accordingly, TQTE and CQTE would be similar, and SQTE would be zero if the relative rank were 0.5, which is not the case in the presented example.

In the following, I illustrate the interpretation of TQTE and SQTE in two important special cases. First, assume that the structural inequality between males and females is zero. This implies that $F_{Y(0)|G}(y|g) = F_{Y(0)}(y)$ for all $y$, and $Q_{Y(0)|G}(\tau|g) = Q_{Y(0)}(\tau)$ for all $\tau$.  
Further, assume that $Y_i(0)$ is continuously distributed such that $F_{Y(0)|G}(Q_{Y(0)|G}(\tau|g)|g) = \tau$. Now, we can show that $\delta_{TQTE}(\tau,g) = \delta_{CQTE}(\tau,g)$
and $\delta_{SQTE}(\tau,g) = 0$ because $\tau_{g}^{r}=\tau$.\footnote{The continuity assumption can be relaxed. Without assuming continuity, \cite{at06} show that $Q_{Y(0)|G}(F_{Y(0)|G}(Q_{Y(0)|G}(\tau|g)|g)|g) = Q_{Y(0)|G}(\tau|g)$ \citep[see Lemma A.1 in][]{at06}. Furthermore, the condition $Q_{Y(1)|G}(F_{Y(0)|G}(Q_{Y(0)|G}(\tau|g)|g)|g) = Q_{Y(1)|G}(\tau|g)$ will hold if $F_{Y(0)|G}(y|g)$ is at least as smooth as $F_{Y(1)|G}(y|g)$. This means that the step size to the next higher rank must satisfy the condition $F_{Y(0)|G}(Q_{Y(0)|G}(\tau|g)|g) \leq F_{Y(1)|G}(Q_{Y(1)|G}(\tau|g)|g)$ for all $\tau$. However, the assumption of continuity is more intuitive and appears reasonable in many applications.} Thus, CQTEs have no structural component, and the entire effect heterogeneity can be directly associated with gender; that is, $\delta_{TQTE}(\tau,f) - \delta_{TQTE}(\tau,m)=\delta_{CQTE}(\tau,f) - \delta_{CQTE}(\tau,m)$ and $\delta_{SQTE}(\tau,f) - \delta_{SQTE}(\tau,m)=0$.

Second, assume that the entire effect heterogeneity by gender can be explained by structural earnings differences. This implies that the transformed quantiles are equal for females and males, 
\begin{equation} \label{qe}
Q_{Y(1)|G}(\tau_{f}^{r}|f) = Q_{Y(1)|G}(\tau_{m}^{r}|m),
\end{equation}
because conditional potential quantiles would not be heterogeneous by gender after rank adjustment. If $Y_i(d)$ is continuously distributed such that $F_{Y(d)|G}(Q_{Y(d)|G}(\tau|g)|g) = \tau$, equation (\ref{qe}) can be rearranged as
\begin{equation} \label{eqqq}
\tau_{f}^{r} = F_{Y(1)|G}(Q_{Y(1)|G}(\tau_{m}^{r}|m)|f).
\end{equation}
The law of iterative expectations for a distribution implies that $F_{Y(1)}(y)= Pr(G_i=f)F_{Y(1)|G}(y|f) + Pr(G_i=m)F_{Y(1)|G}(y|m)$. 
Thus, 
\begin{equation*} 
Pr(G_i=f) \cdot \tau_{f}^{r} = F_{Y(1)}(Q_{Y(1)|G}(\tau_{m}^{r}|m)) - Pr(G_i=m) \cdot F_{Y(1)|G}(Q_{Y(1)|G}(\tau_{m}^{r}|m)|m).
\end{equation*}
Further rearrangements using the abovementioned continuity assumption lead to
\begin{equation*} 
Pr(G_i=f)\cdot \tau_{f}^{r}  +Pr(G_i=m)\cdot \tau_{m}^{r} = F_{Y(1)}(Q_{Y(1)|G}( \tau_{m}^{r}|m)).
\end{equation*}
The left side of this equation is exactly the marginal distribution $F_{Y(0)}(y)= Pr(G_i=f)F_{Y(0)|G}(y|f) + Pr(G_i=m)F_{Y(0)|G}(y|m)$ evaluated at $Q_{Y(0)}(y)$. This leads to the equality $\tau = F_{Y(1)}(Q_{Y(1)|G}(\tau_{m}^{r}|m))$ under the above continuity assumption. If $F_{Y(1)}(y)$ is strictly increasing in the interval $(0,1)$ so that $Q_{Y(1)}(F_{Y(1)}(y))=y$, this result implies that 
\begin{equation*}
Q_{Y(1)}(\tau)= Q_{Y(1)|G}(\tau_{g}^{r}|g)\label{elo}
\end{equation*}
under condition (\ref{qe}).\footnote{This quantile-quantile stability condition is strongly related to the rank stability condition $F_{Y(1)|G}(Q_{Y(1)}(\tau)|g)= F_{Y(0)|G}(Q_{Y(0)}(\tau)|g)$, which is used in \cite{bit05} and \cite{fra15} as a falsification test for rank preservation. Essentially, the researchers' findings imply that the assumption of rank preservation must be rejected if the quantile-quantile stability condition does not hold. TQTEs implicitly test this condition. Although the results of Section \ref{sec6} cannot reject this condition, it is possible that variables other than gender are associated with rank reorganization.} When this finding is substituted into equation (\ref{eq1}), the results are $\delta_{TQTE}(\tau,g)=\delta_{QTE}(\tau)$ and $\delta_{TQTE}(\tau,f) - \delta_{TQTE}(\tau,m) = 0$. Furthermore, $\delta_{SQTE}(\tau,g) = \delta_{CQTE}(\tau,g) - \delta_{QTE}(\tau)$, implying that the entire inter-gender CQTE difference is attributed to the structural effect $\delta_{SQTE}(\tau,f) - \delta_{SQTE}(\tau,m) = \delta_{CQTE}(\tau,f) - \delta_{CQTE}(\tau,m)$.

In summary, TQTE equals CQTE if the structural earnings difference is zero. In contrast, TQTE equals QTE if the heterogeneity between CQTE for males and females can be fully explained by structural earnings inequality. This is a meaningful interpretation for TQTE because QTE represents the unconditional quantile effects.

Finally, the translated conditional average treatment effect (TATE) can be shown to be
\begin{equation*}
\delta_{TATE}(g)=  \int_0^1 \delta_{TQTE}(\tau,g) d\tau = E[Q_{Y(1)|G}(F_{Y(0)|G}(Y_i(0)|g)|g)] - E[Y_i(0)],
\end{equation*}
as an analogy to equation (\ref{eq1}), whereas the structural conditional average treatment effect (SATE) can be shown to be
\begin{equation*}
\delta_{SATE}(g)= \int_0^1 \delta_{SQTE}(\tau,g) d\tau = \delta_{CATE}(g)- \delta_{TATE}(g).
\end{equation*}
These parameters enable the average decomposition to account for differences in outcome structures. The interpretations of TATE and SATE are analogous to those of TQTE and SQTE, respectively. If structural earnings inequality equals zero, then TATE equals CATE and SATE equals zero. If the difference between the CATE for males and females can be fully explained by structural earnings inequality, TATE equals ATE, and SATE is the difference between CATE and ATE. 

\subsection{Estimation}

For an extensive study on estimating counterfactual
distributions under various identification strategies and model specifications, see, e.g.,
\cite{ch09} and \cite{fir10}. In this study, I compute quantiles of earnings by treatment status, which identifies the potential earnings quantiles because of the randomized research design. The most challenging estimation steps are the quantile transformations. The first application of quantile transformation in labour economics dates back to at least \cite{juh91}. I follow the estimation approach proposed by \cite{at06}. They replace the potential outcome distributions by empirical distribution functions. Transformed quantiles can be obtained using a plug-in approach applied to the empirical quantiles and ranks of the empirical distribution functions. The average effects are estimated by replacing the expected values with the transformed sample averages that incorporate quantile transformation. 

In particular, I use the quantile regression estimator
\begin{equation*}
\argmin_{\hat{\beta}_{\tau,0},\hat{\beta}_{\tau,1}} \sum_{i =1}^{N} \omega_i  \cdot\rho_{\tau}(Y_i - \hat{\beta}_{\tau,0} - \hat{\beta}_{\tau,1} D_i ),
\end{equation*}
where $\rho_{\tau}(a)=a(\tau- 1\{a \leq 0\})$ is the check function, and $\omega_i$ represents the sampling and non-response weights ($wgt48b$) provided in the NJCS data. This estimator minimizes the weighted absolute deviations. The hats indicate the estimated parameters. Parameters are $\hat{\beta}_{\tau,0} = \hat{Q}_{Y(0)}(\tau)$ and $\hat{\beta}_{\tau,1} = \hat{\delta}_{QTE}(\tau)$.

For the CQTE estimation, I use the model
\begin{equation} \label{eqapp1}
\argmin_{\hat{\alpha}_{\tau,g,0},\hat{\alpha}_{\tau,g,1}} \sum_{i =1}^{N} 1\{G_i=g\}  \cdot  \omega_i  \cdot \rho_{\tau}(Y_i - \hat{\alpha}_{\tau,g,0} - \hat{\alpha}_{\tau,g,1} D_i ),
\end{equation}
where $1\{\cdot\}$ is the indicator function. Under the randomized research design, parameter $\hat{\alpha}_{\tau,g,1} = \hat{\delta}_{CQTE}(\tau,g)$. Subsequently, I estimate $\hat{Q}_{Y(1|G)}(\tau|g)$ by $\hat{Q}_{Y(1|G)}(\tau|g) = \hat{\alpha}_{\tau,g,0}+\hat{\alpha}_{\tau,g,1}$.

The conditional cumulative distribution functions under non-treatment are estimated using the empirical distribution functions
\begin{equation*}
\hat{F}_{Y(0)|G}(y|g) = \displaystyle \frac{1}{\sum_{i=1}^{N}  1\{D_i=0, G_i=g\} \cdot \omega_i } \sum_{i=1}^{N}  1\{Y_i \leq y,D_i=0, G_i=g\}  \cdot \omega_i.
\end{equation*}
Following  \cite{at06}, the relative ranks are estimated using the plug-in approach $\hat{\tau}_g^r = \hat{F}_{Y(0)|G}(\hat{Q}_{Y(0)}(\tau)|g)$. Now, I can obtain TQTE by replacing the rank $\tau$ with the estimated relative rank $\hat{\tau}_g^r$ in (\ref{eqapp1}),
\begin{equation*}
\argmin_{\hat{\gamma}_{\tau,g,0},\hat{\gamma}_{\tau,g,1}} \sum_{i =1}^{N} 1\{G_i=g\}  \cdot  \omega_i  \cdot \rho_{\hat{\tau}_g^r}(Y_i - \hat{\gamma}_{\tau,g,0} - \hat{\gamma}_{\tau,g,1} D_i ).
\end{equation*}
Parameter $\hat{\gamma}_{\tau,g,1} = \hat{\delta}_{TQTE}(\tau,g)$. The SQTEs are estimated by $\hat{\delta}_{SQTE}(\tau,g)=\hat{\delta}_{CQTE}(\tau,g)-\hat{\delta}_{TQTE}(\tau,g)$. I consider only $\hat{\tau}_g^r \in [0.01, 0.99]$ to avoid estimating extreme quantiles. Furthermore, I exclude all ranks below $\hat{F}_{Y(0)}(0)$ because the continuity assumption is violated at the mass point with zero earnings. This excludes quantiles below the 21st percentile.

\cite{at06} show that the resulting estimates are $\sqrt{n}$ pointwise consistent and asymptotically normal.\footnote{The inference is pointwise and not uniformly valid. Accordingly, the empirical results could be affected by the multiple testing problem.} 
This could 
The standard deviations of all parameters are estimated using a nonparametric bootstrap procedure (sampling individual observations with replacement). \cite{cha15} show the validity of the bootstrap approach for this estimation. \cite{mel15} extend the quantile transformation approach to incorporate exogenous control variables, but I do not need to include additional controls because of the randomized research design. I discuss the estimation of average effects as well as the computation of the Kolmogorov--Smirnov test statistic in Online Appendix \ref{appA}.

\section{Empirical results \label{sec6}}

\subsection{QTE and CQTE}

\begin{figure}
\caption{QTEs of the Job Corps on average weekly earnings (in U.S. dollars) in year four after a randomized assignment.} \label{fig3}
\begin{center}
\includegraphics[width=11cm]{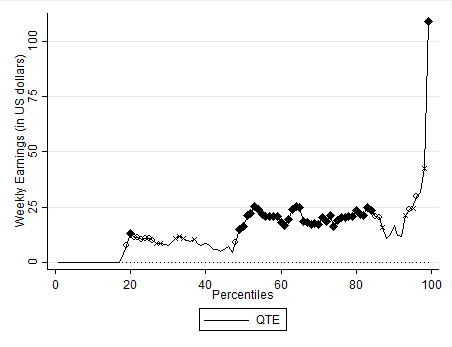}
\end{center}
\parbox{17cm}{\footnotesize Note: The lines report the point estimates calculated separately for all percentiles. Crosses on the lines indicate significant effects at 10\% level. Hollow circles on the lines indicate significant effects at 5\% level. Full diamonds on the lines indicate significant effects at 1\% level.}
\end{figure}

Figure \ref{fig3} shows the QTE of the Job Corps on average weekly earnings four years after the randomized assignment for all percentiles.\footnote{Figure \ref{fig1} in Online Appendix \ref{app_new} shows the potential earnings distributions.} Additionally, column 1 of Table \ref{tab3} documents the deciles of the QTE, including standard errors. The empirical results suggest that quantile differences increase with rank. The QTEs below the 2nd decile do not differ from zero because of participants who are unemployed with no earnings. Between the 2nd and 5th deciles, the earnings gain is approximately 10 U.S. dollars. The earnings gain further increases to approximately 20 U.S. dollars between the 5th and 9th deciles. Beyond the 9th decile, the quantile earnings differences reach values exceeding 100 U.S. dollars. The Kolmogorov--Smirnov test results at the bottom of Table \ref{tab3} support the findings that the Job Corps has a significant positive effect on earnings distribution.

\begin{table}[]
\caption{QTE and CQTE of the Job Corps programme by gender on average weekly earnings (in U.S. dollars) in year four after a randomized assignment.} \label{tab3} \footnotesize
\begin{center}
\begin{tabular}{lcccc}
\hline
\hline
 Quantile          & \multicolumn{ 1}{c}{QTE} & \multicolumn{ 2}{c}{CQTE} & \multicolumn{ 1}{c}{Difference between } \\

           & \multicolumn{ 1}{c}{} &    Females &      Males & \multicolumn{ 1}{c}{(2) and (3)} \\
\cline{2-5}
           &        (1) &        (2) &        (3) &        (4) \\
\hline
0.2 &     13.01*** &      4.425 &      8.011 &     -3.585 \\

           &      (4.166) &      (3.326) &      (8.133) &      (8.730) \\

0.3 &      7.555 &     16.68*** &      2.743 &     13.94 \\

           &      (5.943) &     (6.134) &      (8.735) &     (10.79) \\

0.4 &      8.414 &     19.12** &      7.055 &     12.06 \\

           &      (6.356) &      (8.163) &      (9.690) &     (12.84) \\

0.5 &     16.21*** &      9.059 &     25.08*** &    -16.03 \\

           &      (4.490) &      (9.225) &      (7.387) &     (11.70) \\

0.6 &     17.93*** &     11.99** &     17.25*** &     -5.261 \\

           &      (4.833) &      (6.218) &      (7.749) &      (9.929) \\

0.7 &     17.14*** &     22.36*** &     23.43*** &     -1.070 \\

           &      (4.950) &      (6.536) &      (3.966) &      (7.605) \\

0.8 &     23.28*** &     12.08* &     22.64*** &    -10.56 \\

           &      (6.424) &      (8.259) &      (8.334) &     (11.70) \\

0.9 &     16.51 &     22.38* &     10.03  &     12.35 \\

           &     (11.01) &     (12.38) &     (11.47) &     (16.70) \\

\hline
KS statistic &      11,231*** &       3,610 &      16,029*** &      12,419* \\

           &      [0.002] &      [0.508] &      [0.004] &      [0.086] \\

PSD statistic &      11,231*** &       3,610 &      16,029*** &       1,507 \\

           &      [0.002] &      [0.259] &      [0.001] &      [0.811] \\

NSD statistic &          0 &          0 &          0 &     -12,419** \\

           &      [0.983] &      [0.978] &      [0.991] &      [0.033] \\
\hline
\hline
\end{tabular}   
\end{center}
\parbox{17cm}{\footnotesize Note: *** indicates significance at 1\% level, ** indicates significance at 5\% level, and * indicates significance at 10\% level. Bootstrapped standard errors are in parentheses and are obtained from 1,999 bootstrap replications. ``KS statistic'' denotes the Kolmogorov--Smirnov statistic. ``PSD statistic'' represents the Kolmogorov--Smirnov-type test statistic for positive stochastic dominance. ``NSD statistic'' represents the Kolmogorov--Smirnov-type test statistic for negative stochastic dominance. Bootstrapped p-values (obtained using 1,999 replications) are in square brackets.}
\end{table}

\begin{figure}
\caption{CQTE of the Job Corps programme by gender on average weekly earnings (in U.S. dollars) in year four after a randomized assignment.} \label{fig22}
\begin{center}
\includegraphics[width=11cm]{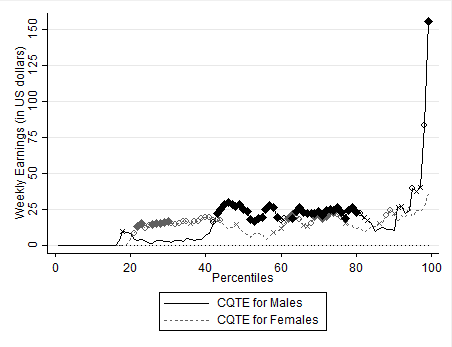}
\end{center}
\parbox{17cm}{\footnotesize Note: The lines represent point estimates calculated separately for all percentiles. Crosses on the lines indicate significant effects at 10\% level. Hollow circles on the lines indicate significant effects at 5\% level. Full diamonds on the lines indicate significant effects at 1\% level.}
\end{figure}

\begin{figure}
\caption{Difference between the Job Corps CQTEs for females and males.} \label{fig4}
\begin{center}
\includegraphics[width=11cm]{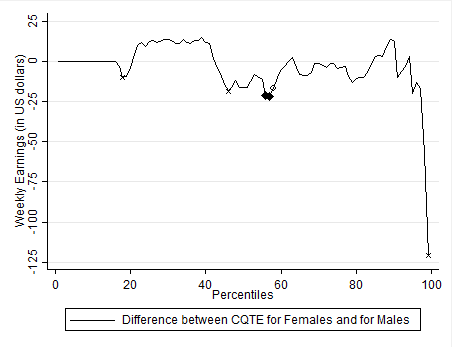}
\end{center}
\parbox{17cm}{\footnotesize Note: The outcome is average weekly earnings (in U.S. dollars) in year four after a randomized assignment. The lines represent the point estimates calculated separately for all percentiles. Crosses on the lines indicate significant effects at 10\% level. Hollow circles on the lines indicate significant effects at 5\% level. Full diamonds on the lines indicate significant effects at 1\% level.}
\end{figure}

The patterns of CQTE for each gender are somewhat different and are reported in Figure \ref{fig22} and columns 2 and 3 of Table \ref{tab3} (the scale of the y-axis differs between Figures \ref{fig3} and \ref{fig22}). For males, the CQTEs are close to zero until the 4th decile. CQTEs jump to a positive impact on earnings of approximately 25 U.S. dollars above the 4th decile. Beyond the 9th decile, the quantile earnings differences exceed 150 U.S. dollars. For females, the CQTEs are zero until the 2nd decile. At the 2nd decile, the quantile earnings differences are approximately 13 U.S. dollars and increase with rank to approximately 25 U.S. dollars. These results are consistent with the findings of \cite{eren14}. The Kolmogorov--Smirnov test results in Table \ref{tab3} suggest that the Job Corps programme has significant positive effects on the potential earnings distribution of males, but no significant effects on that of females.

Figure \ref{fig4} and column 4 of Table \ref{tab3} show the differences between the CQTEs for females and males. The negative differences at the 56th and 57th percentiles are significant at 1\% level. The negative differences at the 58th percentile are significant at 5\% level. The negative differences at the 18th, 46th, and 99th percentiles are significant at 10\% level. Furthermore, the Kolmogorov--Smirnov statistics in Table \ref{tab3} represent the differences between the CQTEs for females and males, which are significant at 10\% level. The test of negative stochastic dominance suggests that the CQTEs for females are always equal to or smaller than the CQTEs for males (at 5\% level). This suggests that the returns of the offer to participate in the Job Corps are larger for males than for females. Accordingly, the Job Corps offers tend to increase gender earnings inequality at some quantiles. 

\subsection{TQTE and SQTE}

\begin{figure}
\caption{Potential earnings distributions by gender.} \label{fig2}
\begin{center}
\includegraphics[width=11cm]{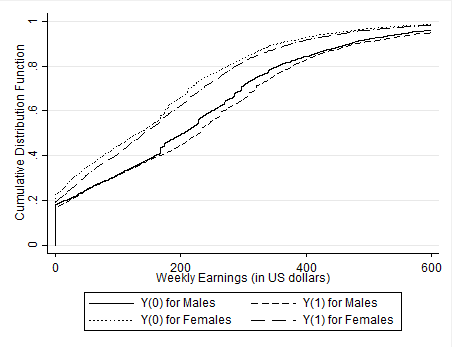}
\end{center}
\parbox{17cm}{\footnotesize Note: The treatment is an offer to join the Job Corps program. The outcome is average weekly earnings (in U.S. dollars) four years after a randomized assignment. The figure is truncated by 600 U.S. dollars in earnings per week.}
\end{figure}

Figure \ref{fig2} shows the potential earnings distributions by gender. Females earn less than males at all quantiles. It is possible that the differences between the CQTEs for females and males can be explained by structural differences between females' and males' earnings distributions.

\begin{figure}
\caption{Rank transformation.} \label{fig34}
\begin{center}
\includegraphics[width=11cm]{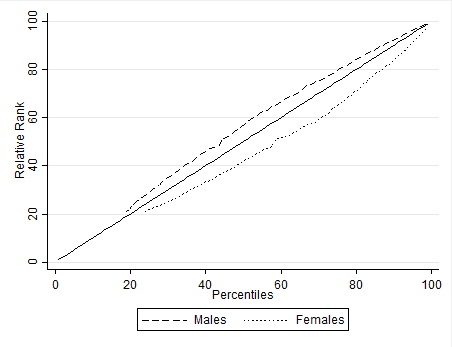}
\end{center}
\parbox{17cm}{\footnotesize Note: I consider the relative ranks only in the interval $[0.01, 0.99]$ and exclude all ranks below $\hat{F}_{Y(0)}(0)$ (for explanation, see Online Appendix \ref{appA}). }
\end{figure}

Figure \ref{fig34} shows the transformed relative ranks on the ordinate and the untransformed ranks on the abscissa. The relative ranks would be on the $45^{\circ}$ line in the absence of structural earnings inequality. The average rank difference between females and males is 12.1 percentage points (the minimum is 2.5, and the maximum is 18.4 percentage points). The relative rank of females is always lower than the $45^{\circ}$ line. The relative rank of males is always higher than the $45^{\circ}$ line. For example, males (females) at the 40th percentile of the conditional potential earnings distribution would have rank 48 (rank 34) in the population distribution. Accordingly, the relative rank shifts the ranks of males upward and the ranks of females downward. This implies that TQTE shifts the CQTE for males to the right and that for females to the left.

\begin{figure}
\caption{TQTEs of the Job Corps by gender on average weekly earnings (in U.S. dollars) in year four after a randomized assignment.} \label{fig5}
\begin{center}
\includegraphics[width=11cm]{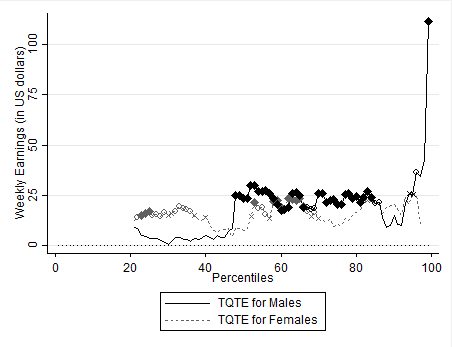}
\end{center}
\parbox{17cm}{\footnotesize Note: The lines represent point estimates calculated separately for all percentiles. Crosses on the lines indicate significant effects at 10\% level. Hollow circles on the lines indicate significant effects at 5\% level. Full diamonds on the lines indicate significant effects at 1\% level. I consider only the relative ranks in the interval $[0.01, 0.99]$ and exclude all ranks below $\hat{F}_{Y(0)}(0)$ (for explanation, see Online Appendix \ref{appA}).}
\end{figure}

\begin{table}[]
\caption{TQTEs and SQTEs of the Job Corps by gender on average weekly earnings (in U.S. dollars) in year four after a randomized assignment.} \label{tab4} \footnotesize
\begin{center}
\begin{tabularx}{17cm}{Xccccccc}
\hline
\hline
 Quantile          & \multicolumn{ 2}{c}{TQTE}& Difference between && \multicolumn{ 2}{c}{SQTE} & Difference between \\

           &    Females &      Males & (1) and (2) &&    Females &      Males & (4) and (5) \\
\cline{2-4} \cline{6-8}
           &        (1) &        (2) &        (3) & &       (4) &        (5) &        (6) \\

\hline
0.3 &     14.83* &      0.650 &     14.13 & &     1.853 &      2.093 &     -0.240 \\

           &      (8.499) &      (8.289) &     (12.107) &  &    (4.431) &      (4.024) &      (6.208) \\

0.4 &     14.18* &      4.863 &      9.314 & &     4.940 &      2.192 &      2.748 \\

           &     (7.994) &      (9.096) &     (12.11) & &     (4.614) &      (4.865) &      (6.795) \\

0.5 &      7.461 &     23.25*** &    -15.79 &   &   1.598 &      1.830 &      -0.232 \\

           &     (6.829) &      (7.846) &     (10.92) &  &    (6.060) &      (6.628) &      (7.034) \\

0.6 &     19.13*** &     17.20*** &      1.928 &  &   -7.133 &      0.056 &     -7.189 \\

           &      (7.391) &      (5.886) &      (8.863) & &     (5.299) &      (4.425) &      (6.587) \\

0.7 &     13.45* &     25.92*** &    -12.47 &  &    8.913* &     -2.485 &     11.40 \\

           &      (7.649) &      (7.326) &    (11.17) &  &    (5.201) &      (5.290) &      (7.737) \\

0.8 &     16.56 &     24.31*** &     -7.747 &  &   -4.479 &     -1.669 &     -2.810 \\

           &     (10.30) &      (7.734) &     (13.11) &    &  (6.720) &     (4.268) &      (7.741) \\

0.9 &     20.30 &     14.79 &      5.501 & &     2.082 &     -4.767 &      6.850 \\

           &     (17.26) &     (12.86) &     (20.56) & &    (12.10) &     (7.161) &     (13.57) \\

\hline
KS statistic &       2,659 &      11,490** &       2,455 & &      1,417 &       4,539 &       2,056 \\

           &      [0.741] &      [0.011] &      [0.909] &  &    [0.988] &      [0.310] &      [0.973] \\

PSD statistic &       2,659 &      11,490*** &       1,610 &  &     1,417 &       4,539 &       1,258 \\

           &      [0.378] &      [0.010] &      [0.727] & &    [0.844] &      [0.112] &      [0.982] \\

NSD statistic &        430 &         67 &      -2,455 & &     -1,361 &      -1,369 &      -2,056 \\

           &      [0.993] &      [0.993] &      [0.551] & &     [0.833] &      [0.904] &      [0.743] \\
\hline
\hline

\end{tabularx}  
\end{center}
\parbox{17cm}{\footnotesize Note: *** indicates significance at 1\% level, ** indicates significance at 5\% level, and * indicates significance at 10\% level. Bootstrapped standard errors are in parentheses and are obtained from 1,999 bootstrap replications. ``KS statistic'' denotes the Kolmogorov--Smirnov statistic. ``PSD statistic'' represents the Kolmogorov--Smirnov-type test statistic for positive stochastic dominance. ``NSD statistic'' represents the Kolmogorov--Smirnov-type test statistic for negative stochastic dominance. Bootstrapped p-values (obtained from 1,999 replications) are in square brackets.}
\end{table}

Figure \ref{fig5} and columns 1 and 2 of Table \ref{tab4} document the findings for TQTEs by gender (the scale of the y-axis differs between Figures \ref{fig22} and \ref{fig5}). The reference distribution is the potential earnings distribution of all eligible Job Corps candidates under non-treatment. This distribution is documented in Figure \ref{fig1} in Online Appendix \ref{app_new}. I do not consider the parts of the potential earnings distribution with zero earnings for quantile transformation because the continuity assumptions---described in Section \ref{sec}---are violated at the mass point with zero earnings. This excludes all ranks lower than the 21st percentile. Furthermore, I exclude the relative ranks higher than the 99th percentile, which avoids estimating extreme quantiles.

Figure \ref{fig5} shows patterns of TQTE that are more consistent between males and females. The TQTE for males is fairly close to zero below the median. Above the median, the quantile earnings differences remain at approximately 25 U.S. dollars. The large difference between the TQTE and CQTE for males at the 99th percentile decreases by approximately 30\% (from 156 to 112 U.S. dollars). The CQTE and TQTE for females are overall quite similar. Below the median, TQTE values are approximately 15 U.S. dollars. The increase to 25 U.S. dollars occurs earlier for TQTE than for CQTE. The Kolmogorov--Smirnov-type test results at the bottom of Table \ref{tab4} show that the Job Corps influences the potential earnings distribution of males, but not that of females. These findings are consistent with the Kolmogorov--Smirnov-type tests applied to CQTE.

\begin{figure}
\caption{Difference between Job Corps TQTEs for females and males.} \label{fig6}
\begin{center}
\includegraphics[width=11cm]{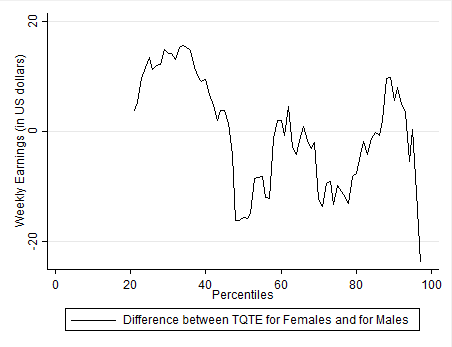}
\end{center}
\parbox{17cm}{\footnotesize Note: The outcome is average weekly earnings (in U.S. dollars) in year four after a randomized assignment. The lines represent the point estimates calculated separately for all percentiles. All point estimates are not statistically different from zero. I consider the relative ranks only in the interval $[0.01, 0.99]$ and exclude all ranks below $\hat{F}_{Y(0)}(0)$ (for explanation, see Online Appendix \ref{appA}).}
\end{figure}

Figure \ref{fig6} and column 3 of Table \ref{tab4} show the quantile differences between the TQTEs for females and males. The differences oscillate around zero. The highly significant negative differences at the 56th and 57th percentiles, which are documented for CQTEs, decline for TQTEs by 45\% (from approximately --22 to --12 U.S. dollars). For TQTEs, I do not observe statistically significant differences between females and males. Tables \ref{tab3} and \ref{tab4} also report standard errors. The standard errors of TQTE are not systematically higher than those of CQTE. Therefore, the insignificant findings are not caused by a comparatively imprecise estimation of TQTE relative to that of CQTE. Furthermore, the bottom rows of Table \ref{tab4} present the Kolmogorov--Smirnov-type statistics. I observe no evidence of difference between the TQTEs of females and males. I also observe no evidence that the TQTEs for females dominate the effects for males or \textit{vice versa}. Overall, these results imply that the direct effects of the Job Corps are not highly heterogeneous by gender.

\begin{figure}
\caption{SQTEs of the Job Corps by gender on average weekly earnings (in U.S. dollars) in year four after a randomized assignment.} \label{fig7}
\begin{center}
\includegraphics[width=11cm]{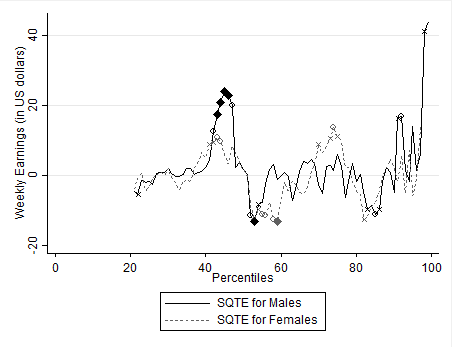}
\end{center}
\parbox{17cm}{\footnotesize Note: The lines represent the point estimates calculated separately for all percentiles. Crosses on the lines indicate significant effects at 10\% level. Hollow circles on the lines indicate significant effects at 5\% level. Full diamonds on the lines indicate significant effects at 1\% level. I consider the relative ranks only in the interval $[0.01, 0.99]$ and exclude all ranks below $\hat{F}_{Y(0)}(0)$ (for an explanation, see Online Appendix \ref{appA}).}
\end{figure}

\begin{figure}
\caption{Difference between SQTEs of the Job Corps for females and males.} \label{fig8}
\begin{center}
\includegraphics[width=11cm]{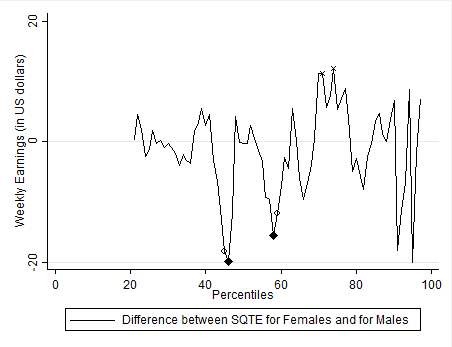}
\end{center}
\parbox{17cm}{\footnotesize Note: The outcome is weekly earnings (in U.S. dollars) in year four after a randomized assignment. The lines represent point estimates calculated separately for all percentiles. Crosses on the lines indicate significant effects at 10\% level. Hollow circles on the lines indicate significant effects at 5\% level. Full diamonds on the lines indicate significant effects at 1\% level. I consider the relative ranks only in the interval $[0.01, 0.99]$ and exclude all ranks below $\hat{F}_{Y(0)}(0)$ (for an explanation, see Online Appendix \ref{appA}).}
\end{figure}

Finally, Figure \ref{fig7} reports the SQTEs by gender, and Figure \ref{fig8} reports the heterogeneity between SQTEs by gender. SQTE is the difference between CQTE and TQTE. Figure \ref{fig7} reports highly significant SQTEs. For males, CQTEs are larger than TQTEs between the 4th and 5th deciles. For females, the positive and negative differences are mixed. Overall, large differences are observed between CQTEs and TQTEs in Figure \ref{fig7}. The differences between SQTEs for males and females are significantly different from zero at some quantiles. Figure \ref{fig8} reports the negative effects at the 46th and 58th percentiles, which are significant at 1\% level, and those at the 45th and 59th percentiles, which are significant at 5\% level. Further, positive effects are reported at the 71st and 74th percentiles, which are significant at 10\% level. Thus, the earnings structure tends to affect CQTE heterogeneity by gender. However, the Kolmogorov--Smirnov-type statistics at the bottom of Table \ref{tab4} do not indicate significant effects. 

In summary, the quantile earnings differences increase with rank. This could favour males because they typically have higher earnings distribution ranks than do females. Accordingly, the Job Corps programme tends to increase the existing inequality in structural earnings. After accounting for expected earnings opportunities, I observe no strong evidence that the Job Corps is more aligned with the training needs of males than those of females.

\subsection{Average effects}

\begin{table}[]
\caption{ATE, CATE, TATE, and SATE of the Job Corps on average weekly earnings (in U.S. dollars) in year four after a randomized assignment.} \label{tab6} \footnotesize
\begin{center}
\begin{tabular}{lcccc}
\hline
\hline
  Gender         & \multicolumn{ 2}{c}{Potential Outcome }  & \multicolumn{ 1}{c}{Average } & Average Differences   \\
					        & \multicolumn{ 2}{c}{ Levels}  & \multicolumn{ 1}{c}{Effects} &  between Females  \\

           & \multicolumn{ 1}{c}{Y(1)} & \multicolumn{ 1}{c}{Y(0)} & \multicolumn{ 1}{c}{} &  and Males \\
\cline{2-5}
           &        (1) &        (2) &        (3) &        (4) \\
\hline
                                      \multicolumn{ 5}{c}{ATE} \\
\hline
  Both Genders         &      210.9 &      195.8 &     15.01*** &    -        \\

           &            &            &      (3.85) &            \\

\hline
                                     \multicolumn{ 5}{c}{CATE} \\
\hline
   Females &      170.4 &      158.1 &     12.31** &     -5.22 \\

           &            &            &      (5.26) &      (7.55) \\

     Males &      239.2 &      221.6 &     17.54*** &      -      \\

           &            &            &      (5.32) &            \\

      \hline
                                     \multicolumn{ 5}{c}{TATE} \\
\hline
   Females &      212.3 &      195.8 &     16.39** &     -0.95 \\

           &            &            &      (6.98) &      (9.22) \\

     Males &      213.6 &      195.8 &     17.34*** &     -       \\

           &            &            &      (5.88) &            \\

\hline
                                     \multicolumn{ 5}{c}{SATE} \\
\hline
   Females &     -       &     -       &     -4.08* &     -4.28 \\

           &            &            &      (2.26) &      (2.96) \\

     Males &    -        &       -     &      0.19 &      -      \\

           &            &            &      (1.70) &            \\

\hline
\hline
\end{tabular}  
\end{center} 
\parbox{17cm}{\footnotesize Note: *** indicates significance at 1\% level, ** indicates significance at 5\% level, and * indicates significance at 10\% level. Bootstrapped standard errors are in parentheses and are obtained from 499 bootstrap replications.}
\end{table}

Table \ref{tab6} shows that individuals awarded offers to participate in the Job Corps experience an average earnings gain of 15 U.S. dollars, or 8\%. Females with no award earn 63 U.S. dollars or 29\% less than males with no award of an offer. In the following discussion, I call this the average gender earnings gap. The offer to participate in the Job Corps increases the earnings of females by 12 U.S. dollars and the earnings of males by 18 U.S. dollars. The average effect is approximately 5 U.S. dollars lower for females than for males, corresponding to approximately 42\% ($=5/12 \cdot 100\%$) higher earnings gain for males than for females. These higher earnings increase the average gender earnings gap by 8\% ($=5/63 \cdot 100\%$). \cite{eren14} and \cite{schoch08} report results of a similar magnitude. In the experimental survey data, the average effect heterogeneity is not significantly different from zero.

TATE accounts for 1 U.S. dollar of the earnings difference between females and males. SATE accounts for 4 U.S. dollars of that difference. Accordingly, the structural earnings inequality could account for up to 82\% and the direct gender effects for only 18\% of the increase in the average gender earnings gap. Despite the insignificant average effects, a large structural earnings inequality can cause the effect heterogeneity by gender. Job Corps trainability differences by gender do not seem to be an important channel of effect heterogeneity. If the offers to participate in the Job Corps are awarded to males and females so that the unconditional potential earnings distributions are aligned under non-treatment, then the average effects are larger for females (average TATE is larger than average CATE), but do not hamper the effects for males. At the same time, gender earnings inequality increases by only 2\% ($=1/63 \cdot 100\%$).

\subsection{Alternative reference distributions and relative ranks}

Online Appendix \ref{appB1} reports TQTE and SQTE when the reference quantiles are from the potential outcome distribution under treatment ($Q_{Y(1)}(\tau)$), the observed outcome distribution ($Q_{Y}(\tau)$), the potential outcome distribution of males under non-treatment ($Q_{Y(0)|G}(\tau|m)$), and the potential outcome distribution of females under non-treatment ($Q_{Y(0)|G}(\tau|f)$). The results do not change qualitatively when a different reference distribution is used (see Figures \ref{figa1}-\ref{figa2}).

Online Appendix \ref{appB2} reports TQTE and SQTE when the relative rank is defined under treatment $\tau_{g}^{r} = F_{Y(1)|G}(Q_{Y_r}(\tau)|g)$. Again, I consider many possible reference distributions (see Figures \ref{figa3}--\ref{figa4}). The results do not change qualitatively when the relative rank is defined under treatment. Accordingly, the results are not sensitive to the choice of the reference distributions and relative ranks.

\section{Effect heterogeneity by gender and parenthood}

\begin{figure}
\caption{Rank transformation.} \label{fig34c}
\begin{center}
\includegraphics[width=11cm]{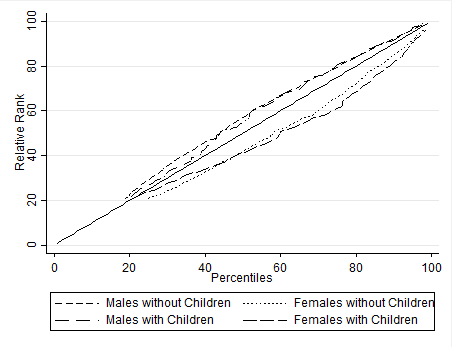}
\end{center}
\parbox{17cm}{\footnotesize Note: I consider the relative ranks only in the interval $[0.01, 0.99]$ and exclude all ranks below $\hat{F}_{Y(0)}(0)$ (for explanation, see Online Appendix \ref{appA}). }
\end{figure}

Having children greatly influences the decision to accept an offer to participate in the Job Corps (see Table \ref{comp}). Therefore, I investigate effect heterogeneity with respect to gender and children. Among the sample's individuals, 33\% are mothers, and 10\% are fathers.\footnote{For 1\% of the sample, the data on children is missing.} Figure \ref{fig34c} shows the relative ranks by gender and parenthood.\footnote{Figure \ref{fig2c} in Online Appendix \ref{app_new} documents the potential earnings distributions under non-treatment by gender and parenthood.} For low percentiles, the relative ranks of mothers and fathers are close to the $45^{\circ}$ line, which indicates little structural gender earnings inequality. However, for high percentiles the relative ranks of mothers are more disperse than for females without children, indicating more structural earnings inequality.

\begin{table}[]
\caption{CATE, TATE, and SATE of the Job Corps by parenthood on average weekly earnings (in U.S. dollars) in year four after a randomized assignment.} \label{tab6c} \footnotesize
\begin{center}
\begin{tabular}{lccccc}
\hline
\hline
  Gender         & \multicolumn{ 2}{c}{With Children }  & \multicolumn{ 2}{c}{Without Children } & Difference between   \\
					      & Level & Difference & Level & Difference & (2) and (4) \\
								\cline{2-6}
							& (1) & (2) & (3) & (4) &(5)	\\

\hline
                                     \multicolumn{ 6}{c}{CATE} \\
\hline
   Females &  20.46***	& 5.40  	& 9.12	& -9.01	 &  14.42 \\

           &   (7.25) &	(17.00)	&	(6.04) &	(8.52) & (21.95)\\

     Males &      15.05	&			&18.13***   &  &  \\

           &          (15.79)		& &	(5.57)	&      &      \\

      \hline
                                     \multicolumn{ 6}{c}{TATE} \\
\hline
   Females &     4.17 &	0.70	 &	-2.49 &	-3.43	& 4.14\\

           &    (5.31) &	(7.43)&		(2.83)&	(2.81) &	(7.86) \\

     Males &    3.47		&  &	0.95 &   & \\

           &      (6.29)  & & (0.65) &  &      \\

\hline
                                     \multicolumn{ 6}{c}{SATE}  \\
\hline
   Females &    16.29***	&4.70		& 11.61* &	-5.58	& 10.28\\

           &       (5.92)	& (14.6)	&	(6.12)	& (8.16)& (22.13) \\

     Males &    11.58	&  & 		17.19***	&   &  \\

           &        	(15.12)		& &	(5.66) &   &         \\

\hline
\hline
\end{tabular}  
\end{center} 
\parbox{17cm}{\footnotesize Note: *** indicates significance at 1\% level, ** indicates significance at 5\% level, and * indicates significance at 10\% level. Bootstrapped standard errors are in parentheses and are obtained from 499 bootstrap replications.}
\end{table}

Table \ref{tab6c} reports suggestive evidence for effect heterogeneity in CATEs by gender and parenthood. On average, mothers benefit more from an offer to join the Job Corps than do fathers. In contrast, childless males profit more from an offer to join the Job Corps than do childless females. Strikingly, the effect of an offer to join the Job Corps is twice as large for mothers as for childless women.\footnote{The average earnings of mothers with an offer to join the Job Corps are 175 U.S. Dollars, and the respective amount is 154 U.S. Dollars for those without an offer. The average earnings of childless women with an offer to join the Job Corps is 169 U.S. Dollars, and the respective amount is 160 U.S. Dollars for those without an offer.} The SATE accounts for 87\% ($=4.7/5.4 \cdot 100\%$) of the CATE heterogeneity by gender for those with children and for 62\% ($=5.58/9.01 \cdot 100\%$) in the absence of children. Even though the effect heterogeneity is insignificant, there is a tendency for the structural earnings inequality to have an economically meaningful influence. Overall, the TATE accounts for 29\% ($=4.14/14.42 \cdot 100\%$) and the SATE for 71\% ($=10.28/14.42 \cdot 100\%$) of the CATE effect heterogeneity by gender and parenthood. These results suggest that within-group heterogeneity in the earnings structures contribute substantially to the between-groups effect heterogeneity. This is consistent with the findings of \cite{ho14}, who document that average effects by group miss important within-group heterogeneity. The results for quantiles are qualitatively similar and are reported in Online Appendix \ref{appE}.

\section{Conclusions}

The existing studies tend to conclude that the Job Corps increases gender earnings inequality. These findings are surprising because most other ALMP evaluations show more positive effects for females than for males \citep[see, e.g.,][]{berg08}. I estimate the TQTEs, initially proposed by \cite{bit13}, to reveal the potential mechanism behind these unexpected findings. \cite{fru12} identify labour market opportunities as the main reason for effect heterogeneity. I approximate such opportunities using expected earnings. Because the earnings distributions differ by gender, the existing structural gender earnings inequality can be a mechanism for effect heterogeneity by gender. 

The results suggest that the award of an offer to participate in the Job Corps increases the average weekly earnings of eligible individuals by approximately 15 U.S. dollars. The absolute earnings gain is approximately 5 U.S. dollars lower for females than for males. The Job Corps offer is least beneficial for childless women. Randomly awarding offers to participate in the Job Corps could increase the gender earnings gap by up to 8\%. My findings suggest that the average structural earnings inequality accounts for 82\% of the effect heterogeneity by gender, whereas only 18\% of the effect heterogeneity can be explained by Job Corps trainability differences by gender. Even though the effect heterogeneity is mostly not significant, this suggests structural earnings differences between groups are important for explaining the overall effect heterogeneity. 

These results suggest that programme administrators should pay more attention to heterogeneous gender earnings structure when designing assignment rules for Job Corps (and potentially also for other ALMPs). An assignment rule that balances the earnings structure of male and female participants would lead to similar effects for both genders. However, several factors have to be considered before implementing such an assignment rule. First, such a rule would crowd-out some males that would significantly benefit from the programme. It is unclear if this is in the interest of the programme administrators. Second, an assignment rule that increases the overall effectiveness of Job Corps would target males and females at the higher end of the earnings distribution. However, such a rule would mainly support those participants who are already well-off. Third, it is unclear how such assignment rules would be implemented in practice. The assignment decision would be based on the potential earnings under non-treatment four years after the programme assignment, which is obviously unobservable at the time the assignment decision is made. Accordingly, the potential earnings under non-treatment would have to be predicted using variables that are observed in the baseline period (e.g., previous earnings, education, etc.), and the assignment rule would have to be based on the predicted potential earnings under non-treatment.\footnote{To illustrate the approach, I predict earnings in the NJCS subsample without an offer to participate in Job Corps with a Tobit model for each gender separately. Table \ref{tobit} in Online Appendix \ref{app_new} documents quite heterogeneous earning prediction models for males and
females eligible for Job Corps. Ethnic origin, drug experience, and past arrests are good earnings predictors for males eligible for Job
Corps. Education and welfare history are good earnings predictors for females eligible for Job Corps. Employment history and living in a PMSA are good earnings predictors for both
genders. The correlation between the predicted and observed earnings is above 30\% for both genders, even though the NJCS contains only relatively 
few covariates.}

There are three potential explanations why many other ALMPs have more positive effects for females than for males. First, other ALMPs could have trainability differences by gender that are in favour of females. Second, many other ALMPs entail non-random selection into treatment, which could also favour females (following the arguments of the preceding paragraph). Third, other ALMPs might mainly support participants with low labour market opportunities, \citep[e.g.,][provide evidence supporting this argument]{card15}, which too could favour females. In the latter case, the structural earnings inequality would also be responsible for effect heterogeneity, but in this case it would support females. However, the influence of structural earnings inequality on the effectiveness of other ALMPs is beyond the scope of this study and is a topic for future research.

\bibliographystyle{econometrica}
\bibliography{Bibliothek_neu}

\clearpage

\renewcommand\appendix{\par
   \setcounter{section}{0}%
   \setcounter{subsection}{0}%
   \setcounter{table}{0}%
	\setcounter{figure}{0}%
   \renewcommand\thesection{\Alph{section}}%
   \renewcommand\thetable{\Alph{section}.\arabic{table}}}
	 \renewcommand\thefigure{\Alph{section}.\arabic{figure}}

\noindent \begin{center}\textbf{\Large -- The following appendices will be made available online. --} \end{center}

\begin{appendix}

\section{Descriptive statistics by gender \label{app_new}}

\begin{figure}[h]
\caption{Potential earnings distribution of all eligible Job Corps candidates.} \label{fig1}
\begin{center}
\includegraphics[width=11cm]{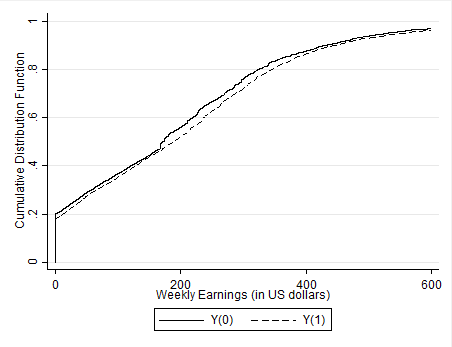}
\end{center}
\parbox{17cm}{\footnotesize Note: The treatment is a offer to join the Job Corps program. The outcome is average weekly earnings (in U.S. dollars) in year four after randomized assignment. The figure is truncated by earnings of 600 U.S. dollars per week.}
\end{figure}

\begin{figure}[h]
\caption{Potential earnings distributions by gender and existence of children.} \label{fig2c}
\begin{center}
\includegraphics[width=11cm]{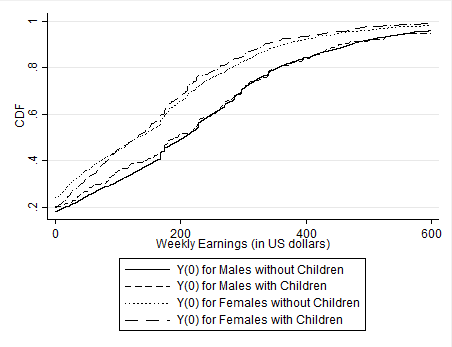}
\end{center}
\parbox{17cm}{\footnotesize Note: The treatment is a offer to join the Job Corps program. The outcome is average weekly earnings (in U.S. dollars) four years after the randomized assignment. The figure is truncated by 600 U.S. dollars in earnings per week.}
\end{figure}

\begin{table}[h]
\caption{Means and standard errors by gender.} \label{tab2} \footnotesize
\begin{center}
\begin{tabularx}{17cm}{Xccccc}
\hline
\hline
           & \multicolumn{ 2}{c}{Males} & \multicolumn{ 2}{c}{Females} & Standardized \\

           &       Mean &  Std. Err. &       Mean &  Std. Err. & Difference \\

           &        (1) &        (2) &        (3) &        (4) &        (5) \\
\hline
Earnings per week in Year 4 &    230.310 &    206.581 &    164.308 &    167.071 &     35.130 \\

Offer to join Job Corps &      0.495 &      0.500 &      0.502 &      0.500 &      1.220 \\

Ever enrolled in a Job Corps center &      0.380 &      0.485 &      0.351 &      0.477 &      6.140 \\
\hline
                  \multicolumn{ 6}{c}{{\bf Socio-economic characteristics}} \\
\hline
Aged between 16-17 years &      0.435 &      0.496 &      0.386 &      0.487 &      9.960 \\

Aged between 18-19 years &      0.310 &      0.463 &      0.327 &      0.469 &      3.620 \\

Aged between 20-24 years &      0.254 &      0.436 &      0.287 &      0.452 &      7.220 \\

     White &      0.305 &      0.460 &      0.222 &      0.415 &     19.010 \\

     Black &      0.450 &      0.497 &      0.519 &      0.500 &     13.950 \\

  Hispanic &      0.169 &      0.375 &      0.185 &      0.388 &      4.180 \\

    Indian &      0.076 &      0.265 &      0.074 &      0.261 &      0.820 \\

Lives with spouse or partner &      0.051 &      0.219 &      0.078 &      0.268 &     11.170 \\

Dummy for bad health &      0.121 &      0.326 &      0.147 &      0.355 &      7.720 \\

0-6 months education program in last year &      0.243 &      0.429 &      0.287 &      0.452 &      9.900 \\

6-12 months education program in last year &      0.385 &      0.486 &      0.324 &      0.468 &     12.660 \\

High school credential &      0.198 &      0.398 &      0.285 &      0.451 &     20.500 \\

Lives in PMSA &      0.316 &      0.465 &      0.334 &      0.471 &      3.770 \\

Lives in MSA &      0.440 &      0.496 &      0.480 &      0.500 &      7.910 \\
\hline
            \multicolumn{ 6}{c}{{\bf Past employment and earnings history}} \\
\hline
Ever had job for two weeks or more &      0.809 &      0.393 &      0.779 &      0.415 &      7.490 \\

Worked in year prior to random assignment &      0.664 &      0.472 &      0.628 &      0.483 &      7.430 \\

Has job at random assignment &      0.212 &      0.409 &      0.203 &      0.402 &      2.240 \\

Below 3 months employed in last year  &      0.190 &      0.378 &      0.190 &      0.382 &      0.080 \\

3-9 months employed in last year  &      0.287 &      0.437 &      0.268 &      0.432 &      4.380 \\

9-12 months employed in last year &      0.187 &      0.377 &      0.171 &      0.366 &      4.360 \\

Yearly earnings less than \$1,000 &      0.099 &      0.299 &      0.123 &      0.328 &      7.590 \\

Yearly earnings \$1,000 to \$5,000 &      0.269 &      0.444 &      0.268 &      0.443 &      0.340 \\

Yearly earnings \$5,000 to \$10,000 &      0.143 &      0.350 &      0.124 &      0.330 &      5.600 \\

Yearly earnings above \$10,000  &      0.079 &      0.269 &      0.050 &      0.219 &     11.530 \\
\hline
                            \multicolumn{ 6}{c}{{\bf Past welfare history}} \\
\hline
Family on welfare when growing up &      0.174 &      0.379 &      0.229 &      0.420 &     13.850 \\

Received food stamps in last year &      0.366 &      0.482 &      0.534 &      0.499 &     34.290 \\

Public or rent-subsidized housing &      0.183 &      0.384 &      0.222 &      0.413 &      9.790 \\

Received AFDC in last year &      0.222 &      0.416 &      0.408 &      0.491 &     40.860 \\
\hline
                                 \multicolumn{ 6}{c}{{\bf Drugs and crime}} \\
\hline
Used hard drugs in last year &      0.076 &      0.265 &      0.050 &      0.218 &     10.620 \\

Smoked marijuana in last year &      0.277 &      0.448 &      0.200 &      0.400 &     18.250 \\

Ever arrested dummy &      0.311 &      0.463 &      0.164 &      0.370 &     35.180 \\
\hline
Observations & \multicolumn{ 2}{c}{6,104} & \multicolumn{ 2}{c}{4,491} &            \\
\hline
\hline
\end{tabularx}
\end{center}
\parbox{17cm}{\footnotesize Note: Time-varying control variables are measured at the time of random assignment. Weights accounting for the sampling design and non-response of the 48-month interview are used ($wgt48b$). PMSA stand for Primary Metropolitan Statistical Area. MSA stand for Metropolitan Statistical Area. AFDC stands for Aid to Families with Dependent Children.}
\end{table}

\begin{table}[]
\caption{Tobit model to predict earnings.} \label{tobit} \footnotesize
\begin{center}
    \begin{tabular}{lcccc}
    \hline  \hline
          & \multicolumn{2}{c}{Males} & \multicolumn{2}{c}{Females} \\
          & Coef. & Std. Err. & Coef. & Std. Err. \\
       \cline{2-5}   & (1)   & (2)   & (3)   & (4) \\
     \hline
    \multicolumn{5}{c}{\textbf{Socio-economic characteristics}} \\
    \hline
    Aged between 18-19 years & 17.8  & (11.9) & -5.96 & (12.9) \\
    Aged between 20-24 years & 2.26  & (15.1) & -16.9 & (16.1) \\
    White & 106*** & (11.4) & -5.54 & (14.5) \\
    Hispanic & 78.4*** & (13.4) & 17.7  & (15.1) \\
    Indian & 78.6*** & (18.1) & 3.84  & (19.7) \\
    Lives with spouse or partner & -9.57 & (23.0) & -19.8 & (21.5) \\
    Dummy for bad health & -12.7 & (14.4) & -11.8 & (14.8) \\
    0-6 months education program in last year & 13.0  & (12.5) & 32.2** & (13.0) \\
    6-12 months education program in last year & 11.6  & (11.8) & 15.4  & (13.4) \\
    High school credential & 21.4  & (13.3) & 51.8*** & (13.4) \\
    Lives in PMSA & 43.6*** & (13.1) & 42.4*** & (15.8) \\
    Lives in MSA & 8.73  & (11.7) & 33.9** & (13.3) \\
    \hline
    \multicolumn{5}{c}{\textbf{Past employment and earnings history}} \\
    \hline
    Ever had job for two weeks or more & 16.1  & (15.4) & 45.5*** & (17.0) \\
    Worked in year prior to random assignment & 65.8** & (25.5) & 10.0  & (29.5) \\
    Has job at random assignment & 8.88  & (11.9) & 28.4* & (14.5) \\
    3-9 months employed in last year  & -7.12 & (16.4) & 19.6  & (18.0) \\
    9-12 months employed in last year & 30.4  & (21.1) & 42.7  & (28.4) \\
    Yearly earnings less than \$1,000 & -21.5 & (25.1) & -8.27 & (29.0) \\
    Yearly earnings \$1,000 to \$5,000 & -15.4 & (21.7) & 27.9  & (25.9) \\
    Yearly earnings \$5,000 to \$10,000 & -1.96 & (23.2) & 22.3  & (29.1) \\
    Yearly earnings above \$10,000  & 37.0  & (29.7) & 82.8** & (35.6) \\
   \hline
    \multicolumn{5}{c}{\textbf{Past welfare history}} \\
   \hline
    Family on welfare when growing up & -3.59 & (13.1) & -30.9*** & (11.9) \\
    Received food stamps in last year & -5.48 & (11.4) & -7.47 & (12.0) \\
    Public or rent-subsidized housing & 0.079 & (13.1) & -16.8 & (11.9) \\
    Received AFDC in last year & -16.6 & (12.4) & 3.42  & (11.9) \\
   \hline
    \multicolumn{5}{c}{\textbf{Drugs and crime}} \\
    \hline
    Used hard drugs in last year & -48.7*** & (18.6) & 20.1  & (29.1) \\
    Smoked marijuana in last year & -11.8 & (10.6) & 3.98  & (12.1) \\
    Ever arrested dummy & -19.4* & (10.4) & -7.24 & (12.9) \\
    \hline
    Observations & \multicolumn{2}{c}{2,620} & \multicolumn{2}{c}{1,603} \\
    \hline \hline
    \end{tabular}%
\end{center} 
\parbox{17cm}{\footnotesize Note: We predict the weekly earnings (in U.S. dollars) 48 months after randomized assignment. *** indicates significance at the 1\% level, ** indicates significance at the 5\% level, and * indicates significance at the 10\% level. Heteroskedastie robust standard errors are in parentheses. }
\end{table}

\clearpage

\section{Estimation \label{appA}}
\subsection{Average effects}

The ATEs are simply estimated using a univariate weighted ordinary least squares (OLS) earnings regression on the treatment dummy. To estimate CATE, I proceed in a similar manner but restrict the sample by gender. The TATEs are computationally very intensive to estimate. Again, I follow the approach of \textbf{at06} and replace the expectations by their sample average. In particular,
\begin{equation*}
\hat{\delta}_{TATE}(g)   = \displaystyle \frac{1}{\sum_{1=1}^{N} (1-D_i) \cdot  \omega_i}  \sum_{1=1}^{N} (1-D_i) \cdot \omega_i \cdot \left(\hat{Q}_{Y(1)|G}(\hat{F}_{Y(0)|G}(Y_i|g) |g) -    Y_i \right).
\end{equation*}
Finally, I estimate SATE by $\hat{\delta}_{SATE}(g)=\hat{\delta}_{CATE}(g)-\hat{\delta}_{TATE}(g)$.

\subsection{Kolmogorov--Smirnov test}

Additionally, I report the rank invariant Kolmogorov--Smirnov test statistics for all effects. The test statistics are
\begin{equation*}
T_{KS} = \sqrt{n} \cdot \sup_{\tau } ( | \hat{\delta}(\tau) | ),
\end{equation*}
where $\hat{\delta}(\tau)$ indicates the estimate of some quantile difference $\delta(\tau)$. The null hypothesis is 
$\delta(\tau)=0$ for all $\tau$. Thus, this is a test for significant effects at some quantile or, in other words, for the inequality between some potential earning distributions. Further, I report signed versions of the Kolmogorov--Smirnov statistics, which enable testing for positive or negative first-order stochastic dominance. Positive stochastic dominance is defined as $\delta(\tau)\geq 0$ for all $\tau$. Accordingly, under (positive) stochastic dominance, $\delta(\tau)$ is never negative regardless of the quantile at which the effect is measured. In contrast, negative stochastic dominance is defined as $\delta(\tau)\leq 0$ for all $\tau$. Accordingly, under negative stochastic dominance, $\delta(\tau)$ is never positive, regardless of the quantile at which the effect is measured.\footnote{In particular, I use the test statistics for positive stochastic dominance $T_{PSD} = \sqrt{n} \cdot \sup_{\tau } ( \hat{\delta}(\tau) )$ and negative stochastic dominance $T_{NSD} = \sqrt{n} \cdot \inf_{\tau } ( \hat{\delta}(\tau) )$.} For all Kolmogorov--Smirnov-type test statistics, I estimate p-values using the recentered bootstrapping approach proposed by \textbf{chern05}. This non-pivotal method computes p-values by avoiding restrictive assumptions. \textbf{chern05} show that recentering can substantially improve the finite-sample power of the Kolmogorov--Smirnov test relative to its uncentered counterpart.




\clearpage

\section{Alternative relative ranks \label{appB}}

\subsection{Relative rank is defined under non-treatment \label{appB1}}

\subsubsection{Reference distribution is the potential outcome distribution under treatment}

\begin{figure}[h]
\caption{TQTEs of the Job Corps by gender on average weekly earnings (in U.S. dollars) in year four after randomized assignment.} \label{figa1}
\begin{center}
\includegraphics[width=11cm]{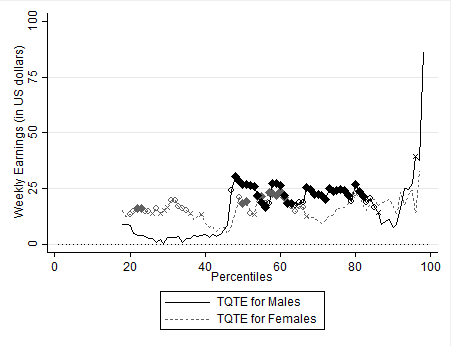}
\end{center}
\parbox{17cm}{\footnotesize Note: The lines report the point estimates separately calculated for all percentiles. Crosses on the lines indicate significant effects at the 10\% level. Hollow circles on the lines indicate significant effects at the 5\% level. Full diamonds on the lines indicate significant effects at the 1\% level. The reference quantiles are from the potential outcome distribution under treatment, $Q_{Y(1)}(\tau)$. The relative rank is defined by $\tau_g^r = F_{Y(0)|G}(Q_{Y(1)}(\tau)|g)$. I consider only the relative ranks in the interval $[0.01, 0.99]$ and exclude all ranks below $\hat{F}_{Y(1)}(0)$ (for explanations, see Online Appendix \ref{appA}).}
\end{figure}

\clearpage

\begin{figure}[h]
\caption{Difference between Job Corps TQTEs for females and males.} \label{fig66}
\begin{center}
\includegraphics[width=11cm]{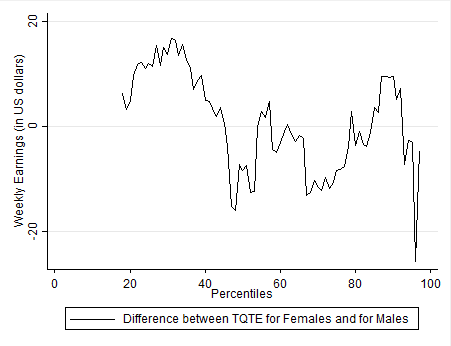}
\end{center}
\parbox{17cm}{\footnotesize Note: The lines report the point estimates calculated separately for all percentiles. All point estimates are not statistically different from zero. The reference quantiles are from the potential outcome distribution under treatment, $Q_{Y(1)}(\tau)$. The relative rank is defined by $\tau_g^r = F_{Y(0)|G}(Q_{Y(1)}(\tau)|g)$. I consider only the relative ranks in the interval $[0.01, 0.99]$ and exclude all ranks below $\hat{F}_{Y(1)}(0)$ (for explanations, see Online Appendix \ref{appA}).}
\end{figure}

\clearpage

\begin{figure}[h]
\caption{SQTEs of the Job Corps by gender on average weekly earnings (in U.S. dollars) in year four after randomized assignment.} 
\begin{center}
\includegraphics[width=11cm]{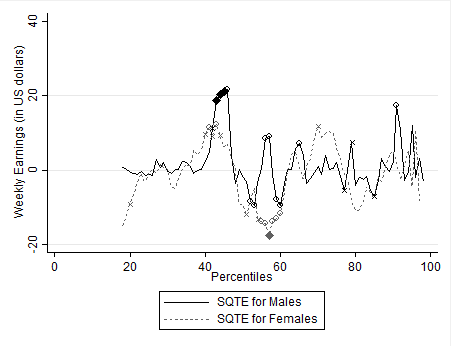}
\end{center}
\parbox{17cm}{\footnotesize Note: The lines report the point estimates separately calculated for all percentiles. Crosses on the lines indicate significant effects at the 10\% level. Hollow circles on the lines indicate significant effects at the 5\% level. Full diamonds on the lines indicate significant effects at the 1\% level. The reference quantiles are from the potential outcome distribution under treatment, $Q_{Y(1)}(\tau)$. The relative rank is defined by $\tau_g^r = F_{Y(0)|G}(Q_{Y(1)}(\tau)|g)$. I consider only the relative ranks in the interval $[0.01, 0.99]$ and exclude all ranks below $\hat{F}_{Y(1)}(0)$ (for explanations, see Online Appendix \ref{appA}).}
\end{figure}

\clearpage

\begin{figure}[h]
\caption{Difference between SQTEs of the Job Corps for females and males.} 
\begin{center}
\includegraphics[width=11cm]{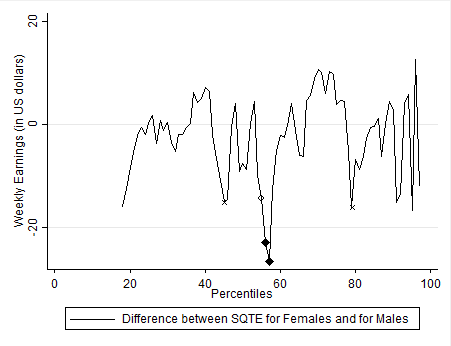}
\end{center}
\parbox{17cm}{\footnotesize Note: The lines report the point estimates calculated separately for all percentiles. Crosses on the lines indicate significant effects at the 10\% level. Hollow circles on the lines indicate significant effects at the 5\% level. Full diamonds on the lines indicate significant effects at the 1\% level. The reference quantiles are from the potential outcome distribution under treatment, $Q_{Y(1)}(\tau)$. The relative rank is defined by $\tau_g^r = F_{Y(0)|G}(Q_{Y(1)}(\tau)|g)$. I consider only the relative ranks in the interval $[0.01, 0.99]$ and exclude all ranks below $\hat{F}_{Y(1)}(0)$ (for explanations, see Online Appendix \ref{appA}).}
\end{figure}

\clearpage
\setcounter{figure}{0}
\subsubsection{Reference distribution is the observed outcome distribution}

\begin{figure}[h]
\caption{TQTEs of the Job Corps by gender on average weekly earnings (in U.S. dollars) in year four after randomized assignment.} 
\begin{center}
\includegraphics[width=11cm]{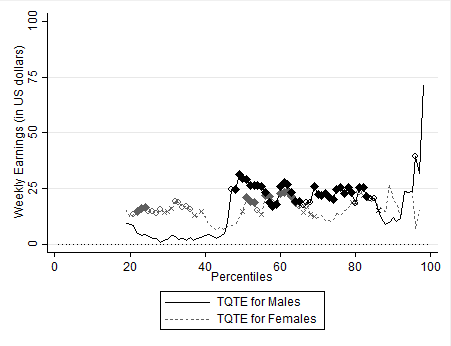}
\end{center}
\parbox{17cm}{\footnotesize Note: The lines report the point estimates calculated separately for all percentiles. Crosses on the lines indicate significant effects at the 10\% level. Hollow circles on the lines indicate significant effects at the 5\% level. Full diamonds on the lines indicate significant effects at the 1\% level. The reference quantiles are from the observed outcome distribution, $Q_{Y}(\tau)$. The relative rank is defined by $\tau_g^r = F_{Y(0)|G}(Q_{Y}(\tau)|g)$. I consider only the relative ranks in the interval $[0.01, 0.99]$ and exclude all ranks below $\hat{F}_{Y}(0)$ (for explanations, see Online Appendix \ref{appA}).}
\end{figure}

\clearpage

\begin{figure}[h]
\caption{Difference between Job Corps TQTEs for females and males.} 
\begin{center}
\includegraphics[width=11cm]{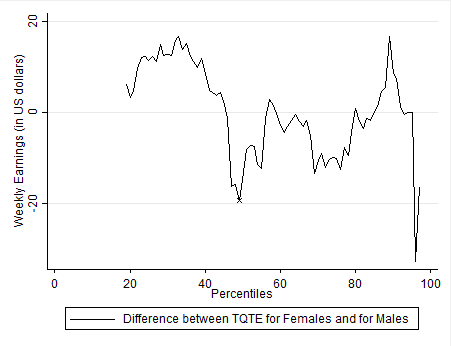}
\end{center}
\parbox{17cm}{\footnotesize Note: The lines report the point estimates calculated separately for all percentiles. Crosses on the lines indicate significant effects at the 10\% level. The reference quantiles are from the observed outcome distribution, $Q_{Y}(\tau)$. The relative rank is defined by $\tau_g^r = F_{Y(0)|G}(Q_{Y}(\tau)|g)$. I consider only the relative ranks in the interval $[0.01, 0.99]$ and exclude all ranks below $\hat{F}_{Y}(0)$ (for explanations, see Online Appendix \ref{appA}).}
\end{figure}

\clearpage

\begin{figure}[h]
\caption{SQTEs of the Job Corps by gender on average weekly earnings (in U.S. dollars) in year four after randomized assignment.} 
\begin{center}
\includegraphics[width=11cm]{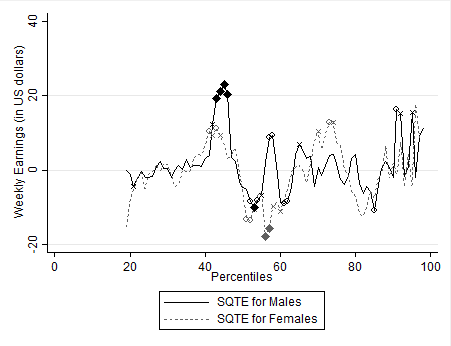}
\end{center}
\parbox{17cm}{\footnotesize Note: The lines report the point estimates calculated separately for all percentiles. Crosses on the lines indicate significant effects at the 10\% level. Hollow circles on the lines indicate significant effects at the 5\% level. Full diamonds on the lines indicate significant effects at the 1\% level. The reference quantiles are from the observed outcome distribution, $Q_{Y}(\tau)$. The relative rank is defined by $\tau_g^r = F_{Y(0)|G}(Q_{Y}(\tau)|g)$. I consider only the relative ranks in the interval $[0.01, 0.99]$ and exclude all ranks below $\hat{F}_{Y}(0)$ (for explanations, see Online Appendix \ref{appA}).}
\end{figure}

\clearpage

\begin{figure}[h]
\caption{Difference between SQTEs of the Job Corps for females and males.} 
\begin{center}
\includegraphics[width=11cm]{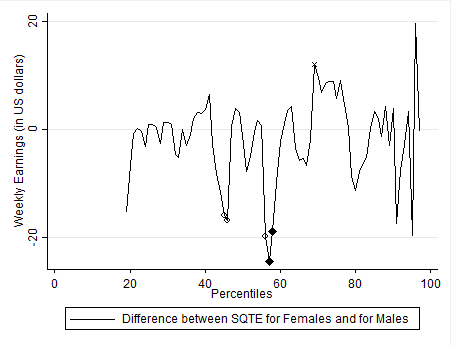}
\end{center}
\parbox{17cm}{\footnotesize Note: The lines report the point estimates calculated separately for all percentiles. Crosses on the lines indicate significant effects at the 10\% level. Hollow circles on the lines indicate significant effects at the 5\% level. Full diamonds on the lines indicate significant effects at the 1\% level. The reference quantiles are from the observed outcome distribution, $Q_{Y}(\tau)$. The relative rank is defined by $\tau_g^r = F_{Y(0)|G}(Q_{Y}(\tau)|g)$. I consider only the relative ranks in the interval $[0.01, 0.99]$ and exclude all ranks below $\hat{F}_{Y}(0)$ (for explanations, see Online Appendix \ref{appA}).}
\end{figure}

\clearpage
\setcounter{figure}{0}
\subsubsection{Reference distribution is the potential outcome distribution of males under non-treatment}

\begin{figure}[h]
\caption{TQTEs of the Job Corps by gender on average weekly earnings (in U.S. dollars) in year four after randomized assignment.} 
\begin{center}
\includegraphics[width=11cm]{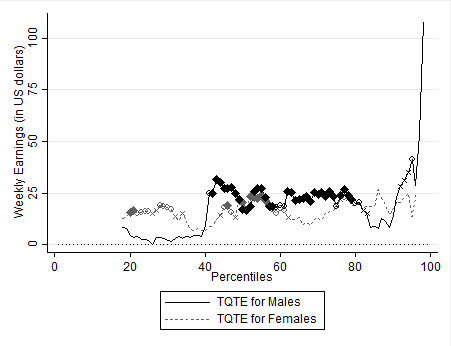}
\end{center}
\parbox{17cm}{\footnotesize Note: The lines report the point estimates calculated separately for all percentiles. Crosses on the lines indicate significant effects at the 10\% level. Hollow circles on the lines indicate significant effects at the 5\% level. Full diamonds on the lines indicate significant effects at the 1\% level. The reference quantiles are from the potential outcome distribution of males under treatment, $Q_{Y(1)|G}(\tau|m)$. The relative rank is defined by $\tau_g^r = F_{Y(0)|G}(Q_{Y(1)|G}(\tau|m)|g)$. I consider only the relative ranks in the interval $[0.01, 0.99]$ and exclude all ranks below $\hat{F}_{Y(1)|G}(0|m)$ (for explanations, see Online Appendix \ref{appA}).}
\end{figure}

\clearpage

\begin{figure}[h]
\caption{Difference between Job Corps TQTEs for females and males.} 
\begin{center}
\includegraphics[width=11cm]{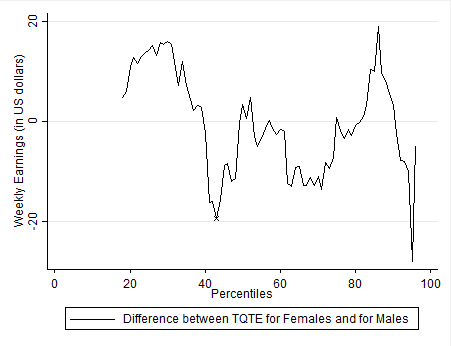}
\end{center}
\parbox{17cm}{\footnotesize Note: The lines report the point estimates calculated separately for all percentiles. Crosses on the lines indicate significant effects at the 10\% level. The reference quantiles are from the potential outcome distribution of males under treatment, $Q_{Y(1)|G}(\tau|m)$. The relative rank is defined by $\tau_g^r = F_{Y(0)|G}(Q_{Y(1)|G}(\tau|m)|g)$. I consider only the relative ranks in the interval $[0.01, 0.99]$ and exclude all ranks below $\hat{F}_{Y(1)|G}(0|m)$ (for explanations, see Online Appendix \ref{appA}).}
\end{figure}

\clearpage

\begin{figure}[h]
\caption{SQTEs of the Job Corps by gender on average weekly earnings (in U.S. dollars) in year four after randomized assignment.} 
\begin{center}
\includegraphics[width=11cm]{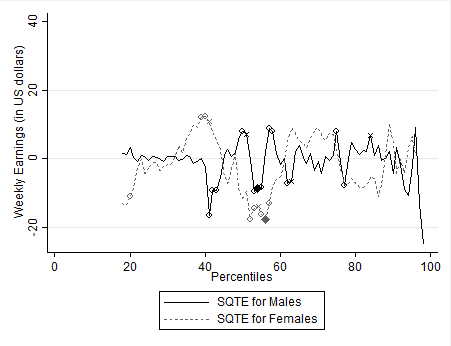}
\end{center}
\parbox{17cm}{\footnotesize Note: The lines report the point estimates calculated separately for all percentiles. Crosses on the lines indicate significant effects at the 10\% level. Hollow circles on the lines indicate significant effects at the 5\% level. Full diamonds on the lines indicate significant effects at the 1\% level. The reference quantiles are from the potential outcome distribution of males under treatment, $Q_{Y(1)|G}(\tau|m)$. The relative rank is defined by $\tau_g^r = F_{Y(0)|G}(Q_{Y(1)|G}(\tau|m)|g)$. I consider only the relative ranks in the interval $[0.01, 0.99]$ and exclude all ranks below $\hat{F}_{Y(1)|G}(0|m)$ (for explanations, see Online Appendix \ref{appA}).}
\end{figure}

\clearpage

\begin{figure}[h]
\caption{Difference between SQTEs of the Job Corps for females and males.} 
\begin{center}
\includegraphics[width=11cm]{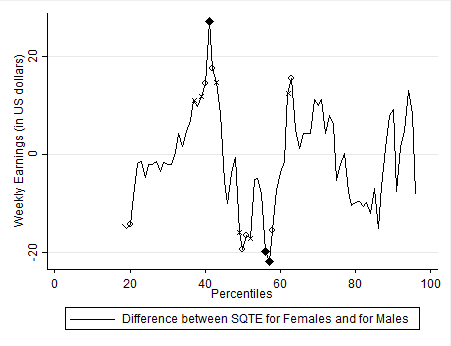}
\end{center}
\parbox{17cm}{\footnotesize Note: The lines report the point estimates calculated separately for all percentiles. Crosses on the lines indicate significant effects at the 10\% level. Hollow circles on the lines indicate significant effects at the 5\% level. Full diamonds on the lines indicate significant effects at the 1\% level. The reference quantiles are from the potential outcome distribution of males under treatment, $Q_{Y(1)|G}(\tau|m)$. The relative rank is defined by $\tau_g^r = F_{Y(0)|G}(Q_{Y(1)|G}(\tau|m)|g)$. I consider only the relative ranks in the interval $[0.01, 0.99]$ and exclude all ranks below $\hat{F}_{Y(1)|G}(0|m)$ (for explanations, see Online Appendix \ref{appA}).}
\end{figure}

\clearpage
\setcounter{figure}{0}
\subsubsection{Reference distribution is the potential outcome distribution of females under non-treatment}

\begin{figure}[h]
\caption{TQTEs of the Job Corps by gender on average weekly earnings (in U.S. dollars) in year four after randomized assignment.} 
\begin{center}
\includegraphics[width=11cm]{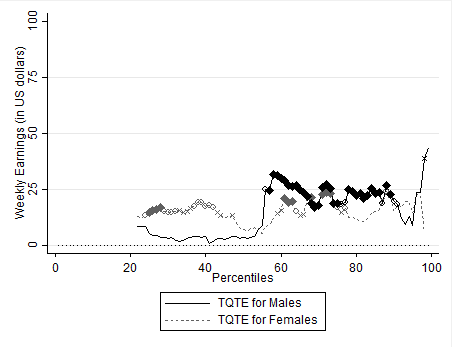}
\end{center}
\parbox{17cm}{\footnotesize Note: The lines report the point estimates calculated separately for all percentiles. Crosses on the lines indicate significant effects at the 10\% level. Hollow circles on the lines indicate significant effects at the 5\% level. Full diamonds on the lines indicate significant effects at the 1\% level. The reference quantiles are from the potential outcome distribution of females under treatment, $Q_{Y(1)|G}(\tau|f)$. The relative rank is defined by $\tau_g^r = F_{Y(0)|G}(Q_{Y(1)|G}(\tau|f)|g)$. I consider only the relative ranks in the interval $[0.01, 0.99]$ and exclude all ranks below $\hat{F}_{Y(1)|G}(0|f)$ (for explanations, see Online Appendix \ref{appA}).}
\end{figure}

\clearpage

\begin{figure}[h]
\caption{Difference between Job Corps TQTEs for females and males.} 
\begin{center}
\includegraphics[width=11cm]{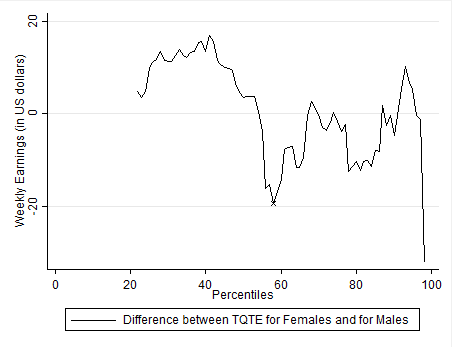}
\end{center}
\parbox{17cm}{\footnotesize Note: The lines report the point estimates calculated separately for all percentiles. Crosses on the lines indicate significant effects at the 10\% level. The reference quantiles are from the potential outcome distribution of females under treatment, $Q_{Y(1)|G}(\tau|f)$. The relative rank is defined by $\tau_g^r = F_{Y(0)|G}(Q_{Y(1)|G}(\tau|f)|g)$. I consider only the relative ranks in the interval $[0.01, 0.99]$ and exclude all ranks below $\hat{F}_{Y(1)|G}(0|f)$ (for explanations, see Online Appendix \ref{appA}).}
\end{figure}

\clearpage

\begin{figure}[h]
\caption{SQTEs of the Job Corps by gender on average weekly earnings (in U.S. dollars) in year four after randomized assignment.} 
\begin{center}
\includegraphics[width=11cm]{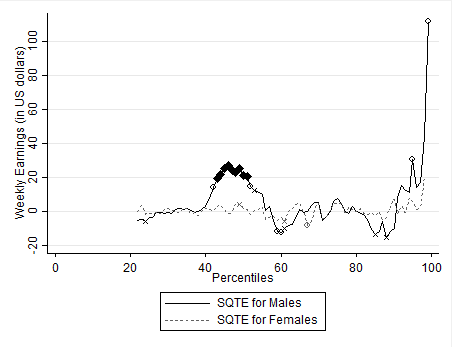}
\end{center}
\parbox{17cm}{\footnotesize Note: The lines report the point estimates calculated separately for all percentiles. Crosses on the lines indicate significant effects at the 10\% level. Hollow circles on the lines indicate significant effects at the 5\% level. Full diamonds on the lines indicate significant effects at the 1\% level. The reference quantiles are from the potential outcome distribution of females under treatment, $Q_{Y(1)|G}(\tau|f)$. The relative rank is defined by $\tau_g^r = F_{Y(0)|G}(Q_{Y(1)|G}(\tau|f)|g)$. I consider only the relative ranks in the interval $[0.01, 0.99]$ and exclude all ranks below $\hat{F}_{Y(1)|G}(0|f)$ (for explanations, see Online Appendix \ref{appA}).}
\end{figure}

\clearpage

\begin{figure}[h]
\caption{Difference between the SQTEs of the Job Corps for females and males.}  \label{figa2}
\begin{center}
\includegraphics[width=11cm]{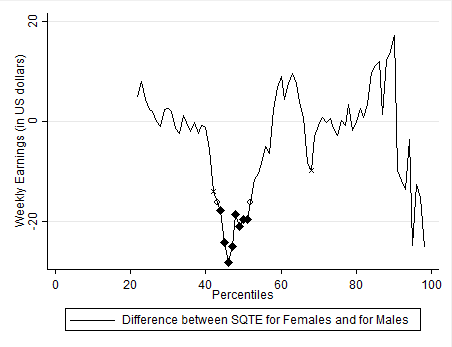}
\end{center}
\parbox{17cm}{\footnotesize Note: The lines report the point estimates calculated separately for all percentiles. Crosses on the lines indicate significant effects at the 10\% level. Hollow circles on the lines indicate significant effects at the 5\% level. Full diamonds on the lines indicate significant effects at the 1\% level. The reference quantiles are from the potential outcome distribution of females under treatment, $Q_{Y(1)|G}(\tau|f)$. The relative rank is defined by $\tau_g^r = F_{Y(0)|G}(Q_{Y(1)|G}(\tau|f)|g)$. I consider only the relative ranks in the interval $[0.01, 0.99]$ and exclude all ranks below $\hat{F}_{Y(1)|G}(0|f)$ (for explanations, see Online Appendix \ref{appA}).}
\end{figure}

\clearpage
\setcounter{figure}{0}
\subsection{Relative rank is defined under treatment \label{appB2}}
\subsubsection{Reference distribution is the potential outcome distribution under non-treatment}

\begin{figure}[h]
\caption{TQTEs of the Job Corps by gender on average weekly earnings (in U.S. dollars) in year four after randomized assignment.} \label{figa3}
\begin{center}
\includegraphics[width=11cm]{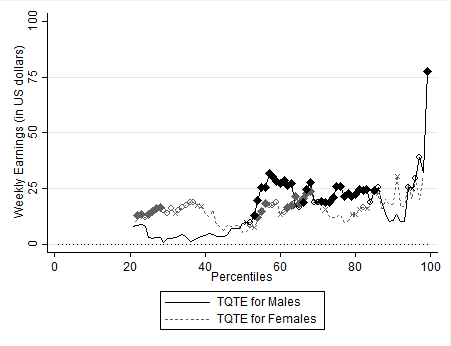}
\end{center}
\parbox{17cm}{\footnotesize Note: The lines report the point estimates calculated separately for all percentiles. Crosses on the lines indicate significant effects at the 10\% level. Hollow circles on the lines indicate significant effects at the 5\% level. Full diamonds on the lines indicate significant effects at the 1\% level. The reference quantiles are from the potential outcome distribution under non-treatment, $Q_{Y(0)}(\tau)$. The relative rank is defined by $\tau_g^r = F_{Y(1)|G}(Q_{Y(0)}(\tau)|g)$. The rank transformation uses the conditional potential outcome distribution under treatment. I consider only the relative ranks in the interval $[0.01, 0.99]$ and exclude all ranks below $\hat{F}_{Y(1)}(0)$ (for explanations, see Online Appendix \ref{appA}).}
\end{figure}

\clearpage

\begin{figure}[h]
\caption{Difference between Job Corps TQTEs for females and males.} 
\begin{center}
\includegraphics[width=11cm]{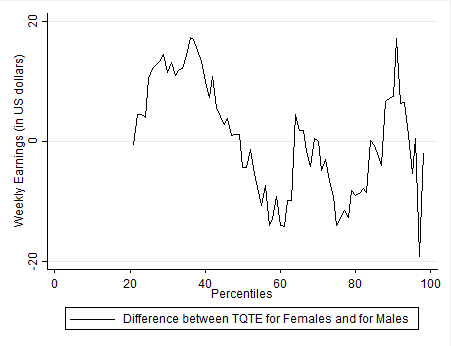}
\end{center}
\parbox{17cm}{\footnotesize Note: The lines report the point estimates calculated separately for all percentiles. All point estimates are not statistically different from zero. The reference quantiles are from the potential outcome distribution under non-treatment, $Q_{Y(0)}(\tau)$. The relative rank is defined by $\tau_g^r = F_{Y(1)|G}(Q_{Y(0)}(\tau)|g)$. The rank transformation uses the conditional potential outcome distribution under treatment. I consider only the relative ranks in the interval $[0.01, 0.99]$ and exclude all ranks below $\hat{F}_{Y(1)}(0)$ (for explanations, see Online Appendix \ref{appA}).}
\end{figure}

\clearpage

\begin{figure}[h]
\caption{SQTEs of the Job Corps by gender on average weekly earnings (in U.S. dollars) in year four after randomized assignment.} 
\begin{center}
\includegraphics[width=11cm]{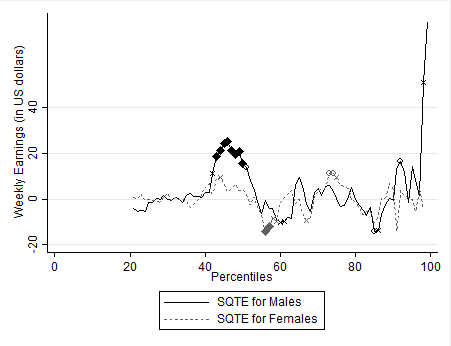}
\end{center}
\parbox{17cm}{\footnotesize Note: The lines report the point estimates calculated separately for all percentiles. Crosses on the lines indicate significant effects at the 10\% level. Hollow circles on the lines indicate significant effects at the 5\% level. Full diamonds on the lines indicate significant effects at the 1\% level. The reference quantiles are from the potential outcome distribution under non-treatment, $Q_{Y(0)}(\tau)$. The relative rank is defined by $\tau_g^r = F_{Y(1)|G}(Q_{Y(0)}(\tau)|g)$. The rank transformation uses the conditional potential outcome distribution under treatment. I consider only the relative ranks in the interval $[0.01, 0.99]$ and exclude all ranks below $\hat{F}_{Y(1)}(0)$ (for explanations, see Online Appendix \ref{appA}).}
\end{figure}

\clearpage

\begin{figure}[h]
\caption{Difference between SQTEs of the Job Corps for females and males.} 
\begin{center}
\includegraphics[width=11cm]{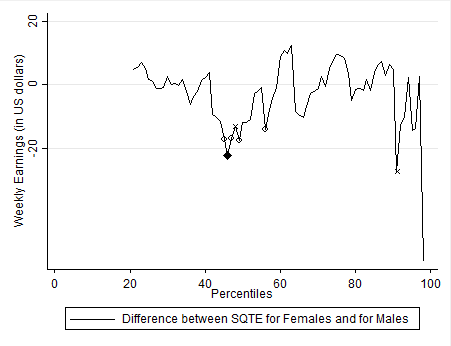}
\end{center}
\parbox{17cm}{\footnotesize Note: The lines report the point estimates calculated separately for all percentiles. Crosses on the lines indicate significant effects at the 10\% level. Hollow circles on the lines indicate significant effects at the 5\% level. Full diamonds on the lines indicate significant effects at the 1\% level. The reference quantiles are from the potential outcome distribution under non-treatment, $Q_{Y(0)}(\tau)$. The relative rank is defined by $\tau_g^r = F_{Y(1)|G}(Q_{Y(0)}(\tau)|g)$. The rank transformation uses the conditional potential outcome distribution under treatment. I consider only the relative ranks in the interval $[0.01, 0.99]$ and exclude all ranks below $\hat{F}_{Y(1)}(0)$ (for explanations, see Online Appendix \ref{appA}).}
\end{figure}

\clearpage
\setcounter{figure}{0}
\subsubsection{Reference distribution is the potential outcome distribution under treatment}

\begin{figure}[h]
\caption{TQTEs of the Job Corps by gender on average weekly earnings (in U.S. dollars) in year four after randomized assignment.} 
\begin{center}
\includegraphics[width=11cm]{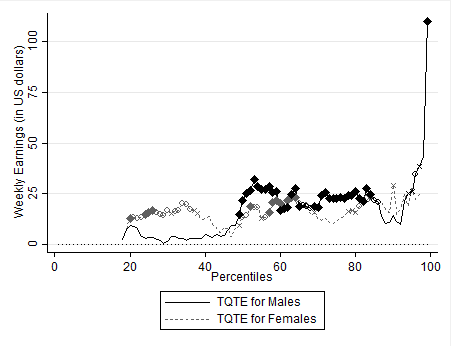}
\end{center}
\parbox{17cm}{\footnotesize Note: The lines report the point estimates calculated separately for all percentiles. Crosses on the lines indicate significant effects at the 10\% level. Hollow circles on the lines indicate significant effects at the 5\% level. Full diamonds on the lines indicate significant effects at the 1\% level. The reference quantiles are from the potential outcome distribution under treatment, $Q_{Y(1)}(\tau)$. The relative rank is defined by $\tau_g^r = F_{Y(1)|G}(Q_{Y(1)}(\tau)|g)$. The rank transformation uses the conditional potential outcome distribution under treatment. I consider only the relative ranks in the interval $[0.01, 0.99]$ and exclude all ranks below $\hat{F}_{Y(1)}(0)$ (for explanations, see Online Appendix \ref{appA}).}
\end{figure}

\clearpage

\begin{figure}[h]
\caption{Difference between Job Corps TQTEs for females and males.}
\begin{center}
\includegraphics[width=11cm]{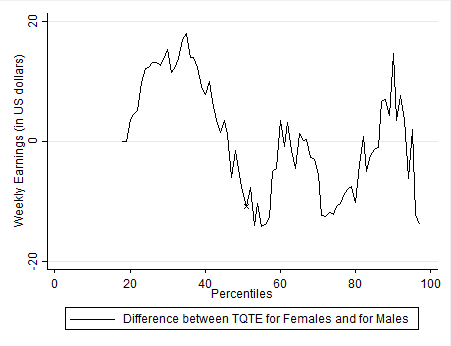}
\end{center}
\parbox{17cm}{\footnotesize Note: The lines report the point estimates calculated separately for all percentiles. Crosses on the lines indicate significant effects at the 10\% level. The reference quantiles are from the potential outcome distribution under treatment, $Q_{Y(1)}(\tau)$. The relative rank is defined by $\tau_g^r = F_{Y(1)|G}(Q_{Y(1)}(\tau)|g)$. The rank transformation uses the conditional potential outcome distribution under treatment. I consider only the relative ranks in the interval $[0.01, 0.99]$ and exclude all ranks below $\hat{F}_{Y(1)}(0)$ (for explanations, see Online Appendix \ref{appA}).}
\end{figure}

\clearpage

\begin{figure}[h]
\caption{SQTEs of the Job Corps by gender on average weekly earnings (in U.S. dollars) in year four after randomized assignment.} 
\begin{center}
\includegraphics[width=11cm]{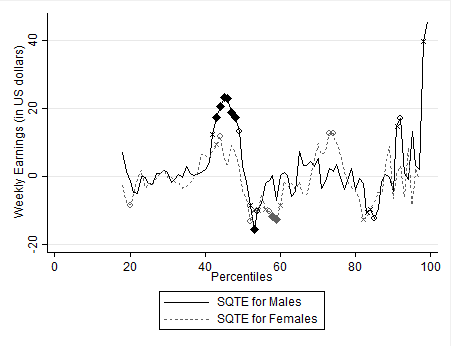}
\end{center}
\parbox{17cm}{\footnotesize Note: The lines report the point estimates calculated separately for all percentiles. Crosses on the lines indicate significant effects at the 10\% level. Hollow circles on the lines indicate significant effects at the 5\% level. Full diamonds on the lines indicate significant effects at the 1\% level. The reference quantiles are from the potential outcome distribution under treatment, $Q_{Y(1)}(\tau)$. The relative rank is defined by $\tau_g^r = F_{Y(1)|G}(Q_{Y(1)}(\tau)|g)$. The rank transformation uses the conditional potential outcome distribution under treatment. I consider only the relative ranks in the interval $[0.01, 0.99]$ and exclude all ranks below $\hat{F}_{Y(1)}(0)$ (for explanations, see Online Appendix \ref{appA}).}
\end{figure}

\clearpage

\begin{figure}[h]
\caption{Difference between SQTEs of the Job Corps for females and males.} 
\begin{center}
\includegraphics[width=11cm]{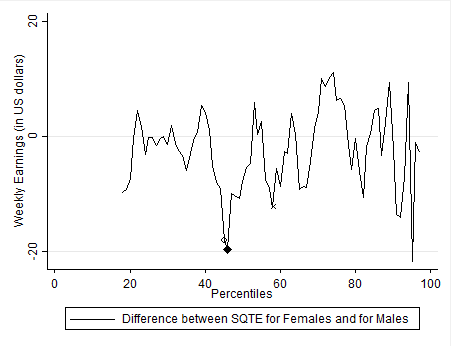}
\end{center}
\parbox{17cm}{\footnotesize Note: The lines report the point estimates calculated separately for all percentiles. Crosses on the lines indicate significant effects at the 10\% level. Hollow circles on the lines indicate significant effects at the 5\% level. Full diamonds on the lines indicate significant effects at the 1\% level. The reference quantiles are from the potential outcome distribution under treatment, $Q_{Y(1)}(\tau)$. The relative rank is defined by $\tau_g^r = F_{Y(1)|G}(Q_{Y(1)}(\tau)|g)$. The rank transformation uses the conditional potential outcome distribution under treatment. I consider only the relative ranks in the interval $[0.01, 0.99]$ and exclude all ranks below $\hat{F}_{Y(1)}(0)$ (for explanations, see Online Appendix \ref{appA}).}
\end{figure}

\clearpage \setcounter{figure}{0}
\subsubsection{Reference distribution is the observed outcome distribution}

\begin{figure}[h]
\caption{TQTEs of the Job Corps by gender on average weekly earnings (in U.S. dollars) in year four after randomized assignment.} 
\begin{center}
\includegraphics[width=11cm]{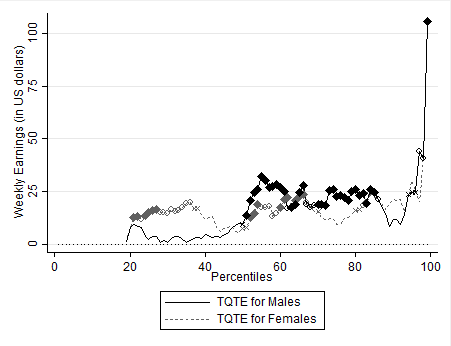}
\end{center}
\parbox{17cm}{\footnotesize Note: The lines report the point estimates calculated separately for all percentiles. Crosses on the lines indicate significant effects at the 10\% level. Hollow circles on the lines indicate significant effects at the 5\% level. Full diamonds on the lines indicate significant effects at the 1\% level. The reference quantiles are from the observed outcome distribution, $Q_{Y}(\tau)$. The relative rank is defined by $\tau_g^r = F_{Y(1)|G}(Q_{Y}(\tau)|g)$. The rank transformation uses the conditional potential outcome distribution under treatment. I consider only the relative ranks in the interval $[0.01, 0.99]$ and exclude all ranks below $\hat{F}_{Y}(0)$ (for explanations, see Online Appendix \ref{appA}).}
\end{figure}

\clearpage

\begin{figure}[h]
\caption{Difference between Job Corps TQTEs for females and males.} 
\begin{center}
\includegraphics[width=11cm]{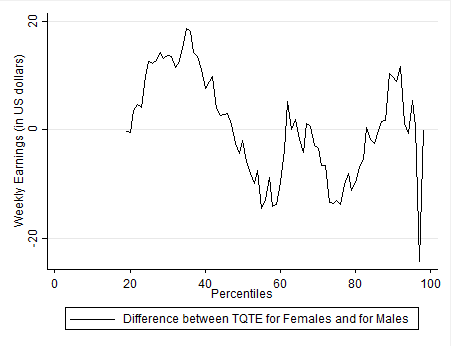}
\end{center}
\parbox{17cm}{\footnotesize Note: The lines report the point estimates calculated separately for all percentiles. All point estimates are not statistically different from zero. The reference quantiles are from the observed outcome distribution, $Q_{Y}(\tau)$. The relative rank is defined by $\tau_g^r = F_{Y(1)|G}(Q_{Y}(\tau)|g)$. The rank transformation uses the conditional potential outcome distribution under treatment. I consider only the relative ranks in the interval $[0.01, 0.99]$ and exclude all ranks below $\hat{F}_{Y}(0)$ (for explanations, see Online Appendix \ref{appA}).}
\end{figure}

\clearpage

\begin{figure}[h]
\caption{SQTEs of the Job Corps by gender on average weekly earnings (in U.S. dollars) in year four after randomized assignment.} 
\begin{center}
\includegraphics[width=11cm]{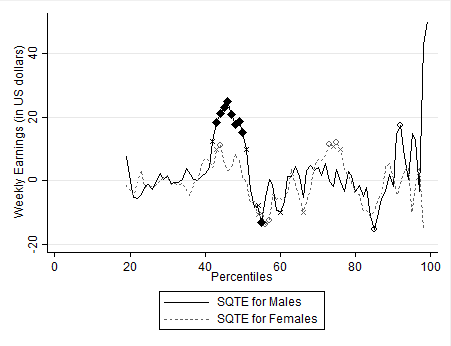}
\end{center}
\parbox{17cm}{\footnotesize Note: The lines report the point estimates calculated separately for all percentiles. Crosses on the lines indicate significant effects at the 10\% level. Hollow circles on the lines indicate significant effects at the 5\% level. Full diamonds on the lines indicate significant effects at the 1\% level. The reference quantiles are from the observed outcome distribution, $Q_{Y}(\tau)$. The relative rank is defined by $\tau_g^r = F_{Y(1)|G}(Q_{Y}(\tau)|g)$. The rank transformation uses the conditional potential outcome distribution under treatment. I consider only the relative ranks in the interval $[0.01, 0.99]$ and exclude all ranks below $\hat{F}_{Y}(0)$ (for explanations, see Online Appendix \ref{appA}).}
\end{figure}

\clearpage

\begin{figure}[h]
\caption{Difference between SQTEs of the Job Corps for females and males.} 
\begin{center}
\includegraphics[width=11cm]{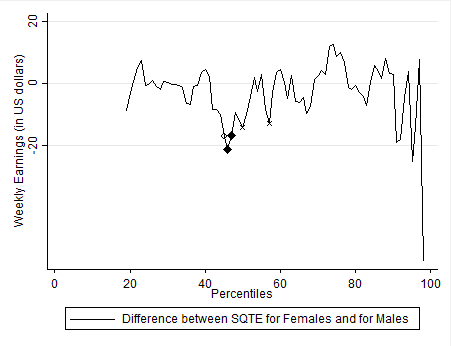}
\end{center}
\parbox{17cm}{\footnotesize Note: The lines report the point estimates calculated separately for all percentiles. Crosses on the lines indicate significant effects at the 10\% level. Hollow circles on the lines indicate significant effects at the 5\% level. Full diamonds on the lines indicate significant effects at the 1\% level. The reference quantiles are from the observed outcome distribution, $Q_{Y}(\tau)$. The relative rank is defined by $\tau_g^r = F_{Y(1)|G}(Q_{Y}(\tau)|g)$. The rank transformation uses the conditional potential outcome distribution under treatment. I consider only the relative ranks in the interval $[0.01, 0.99]$ and exclude all ranks below $\hat{F}_{Y}(0)$ (for explanations, see Online Appendix \ref{appA}).}
\end{figure}

\clearpage \setcounter{figure}{0}
\subsubsection{Reference distribution is the potential outcome distribution of males under non-treatment}

\begin{figure}[h]
\caption{TQTEs of the Job Corps by gender on average weekly earnings (in U.S. dollars) in year four after randomized assignment.} 
\begin{center}
\includegraphics[width=11cm]{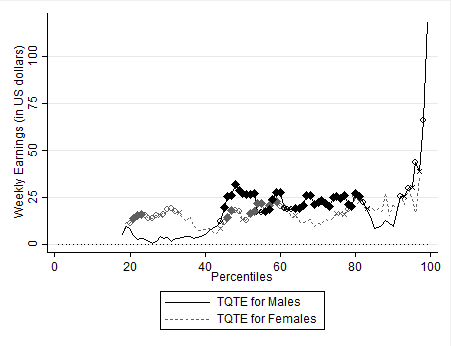}
\end{center}
\parbox{17cm}{\footnotesize Note: The lines report the point estimates calculated separately for all percentiles. Crosses on the lines indicate significant effects at the 10\% level. Hollow circles on the lines indicate significant effects at the 5\% level. Full diamonds on the lines indicate significant effects at the 1\% level. The reference quantiles are from the potential outcome distribution of males under treatment, $Q_{Y(0)|G}(\tau|m)$. The relative rank is defined by $\tau_g^r = F_{Y(1)|G}(Q_{Y(0)|G}(\tau|m)|g)$. The rank transformation uses the conditional potential outcome distribution under treatment. I consider only the relative ranks in the interval $[0.01, 0.99]$ and exclude all ranks below $\hat{F}_{Y(0)|G}(0|m)$ (for explanations, see Online Appendix \ref{appA}).}
\end{figure}

\clearpage

\begin{figure}[h]
\caption{Difference between Job Corps TQTEs for females and males.} 
\begin{center}
\includegraphics[width=11cm]{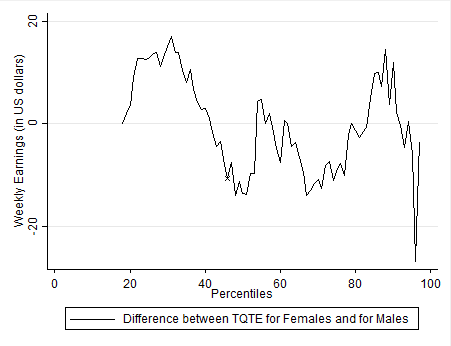}
\end{center}
\parbox{17cm}{\footnotesize Note: The lines report the point estimates calculated separately for all percentiles. Crosses on the lines indicate significant effects at the 10\% level. The reference quantiles are from the potential outcome distribution of males under treatment, $Q_{Y(0)|G}(\tau|m)$. The relative rank is defined by $\tau_g^r = F_{Y(1)|G}(Q_{Y(0)|G}(\tau|m)|g)$. The rank transformation uses the conditional potential outcome distribution under treatment. I consider only the relative ranks in the interval $[0.01, 0.99]$ and exclude all ranks below $\hat{F}_{Y(0)|G}(0|m)$ (for explanations, see Online Appendix \ref{appA}).}
\end{figure}

\clearpage

\begin{figure}[h]
\caption{SQTEs of the Job Corps by gender on average weekly earnings (in U.S. dollars) in year four after randomized assignment.} 
\begin{center}
\includegraphics[width=11cm]{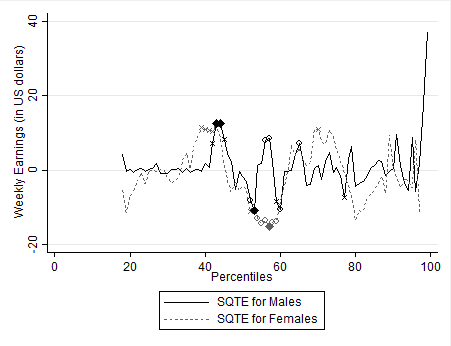}
\end{center}
\parbox{17cm}{\footnotesize Note: The lines report the point estimates calculated separately for all percentiles. Crosses on the lines indicate significant effects at the 10\% level. Hollow circles on the lines indicate significant effects at the 5\% level. Full diamonds on the lines indicate significant effects at the 1\% level. The reference quantiles are from the potential outcome distribution of males under treatment, $Q_{Y(0)|G}(\tau|m)$. The relative rank is defined by $\tau_g^r = F_{Y(1)|G}(Q_{Y(0)|G}(\tau|m)|g)$. The rank transformation uses the conditional potential outcome distribution under treatment. I consider only the relative ranks in the interval $[0.01, 0.99]$ and exclude all ranks below $\hat{F}_{Y(0)|G}(0|m)$ (for explanations, see Online Appendix \ref{appA}).}
\end{figure}

\clearpage

\begin{figure}[h]
\caption{Difference between SQTEs of the Job Corps for females and males.} 
\begin{center}
\includegraphics[width=11cm]{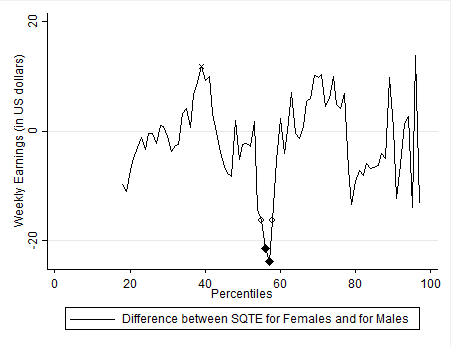}
\end{center}
\parbox{17cm}{\footnotesize Note: The lines report the point estimates calculated separately for all percentiles. Crosses on the lines indicate significant effects at the 10\% level. Hollow circles on the lines indicate significant effects at the 5\% level. Full diamonds on the lines indicate significant effects at the 1\% level. The reference quantiles are from the potential outcome distribution of males under treatment, $Q_{Y(0)|G}(\tau|m)$. The relative rank is defined by $\tau_g^r = F_{Y(1)|G}(Q_{Y(0)|G}(\tau|m)|g)$. The rank transformation uses the conditional potential outcome distribution under treatment. I consider only the relative ranks in the interval $[0.01, 0.99]$ and exclude all ranks below $\hat{F}_{Y(0)|G}(0|m)$ (for explanations, see Online Appendix \ref{appA}).}
\end{figure}

\clearpage \setcounter{figure}{0}
\subsubsection{Reference distribution is the potential outcome distribution of females under non-treatment}

\begin{figure}[h]
\caption{TQTEs of the Job Corps by gender on average weekly earnings (in U.S. dollars) in year four after randomized assignment.} 
\begin{center}
\includegraphics[width=11cm]{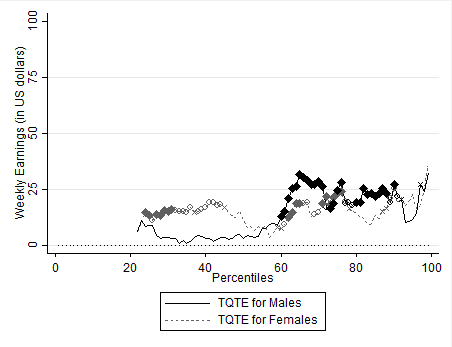}
\end{center}
\parbox{17cm}{\footnotesize Note: The lines report the point estimates calculated separately for all percentiles. Crosses on the lines indicate significant effects at the 10\% level. Hollow circles on the lines indicate significant effects at the 5\% level. Full diamonds on the lines indicate significant effects at the 1\% level. The reference quantiles are from the potential outcome distribution of females under treatment, $Q_{Y(0)|G}(\tau|f)$. The relative rank is defined by $\tau_g^r = F_{Y(1)|G}(Q_{Y(0)|G}(\tau|f)|g)$. The rank transformation uses the conditional potential outcome distribution under treatment. I consider only the relative ranks in the interval $[0.01, 0.99]$ and exclude all ranks below $\hat{F}_{Y(0)|G}(0|f)$ (for explanations, see Online Appendix \ref{appA}).}
\end{figure}

\clearpage

\begin{figure}[h]
\caption{Difference between Job Corps TQTEs for females and males.} 
\begin{center}
\includegraphics[width=11cm]{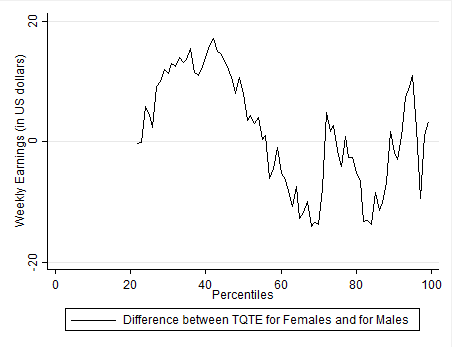}
\end{center}
\parbox{17cm}{\footnotesize Note: The lines report the point estimates calculated separately for all percentiles. All point estimates are not statistically different from zero. The reference quantiles are from the potential outcome distribution of females under treatment, $Q_{Y(0)|G}(\tau|f)$. The relative rank is defined by $\tau_g^r = F_{Y(1)|G}(Q_{Y(0)|G}(\tau|f)|g)$. The rank transformation uses the conditional potential outcome distribution under treatment. I consider only the relative ranks in the interval $[0.01, 0.99]$ and exclude all ranks below $\hat{F}_{Y(0)|G}(0|f)$ (for explanations, see Online Appendix \ref{appA}).}
\end{figure}

\clearpage

\begin{figure}[h]
\caption{SQTEs of the Job Corps by gender on average weekly earnings (in U.S. dollars) in year four after randomized assignment.} 
\begin{center}
\includegraphics[width=11cm]{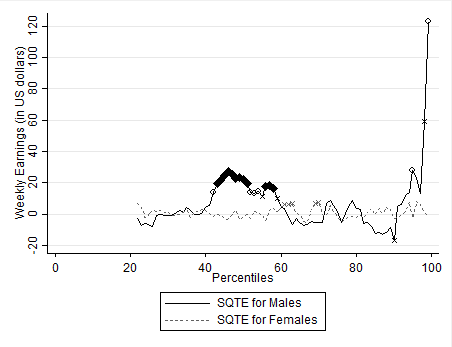}
\end{center}
\parbox{17cm}{\footnotesize Note: The lines report the point estimates calculated separately for all percentiles. Crosses on the lines indicate significant effects at the 10\% level. Hollow circles on the lines indicate significant effects at the 5\% level. Full diamonds on the lines indicate significant effects at the 1\% level. The reference quantiles are from the potential outcome distribution of females under treatment, $Q_{Y(0)|G}(\tau|f)$. The relative rank is defined by $\tau_g^r = F_{Y(1)|G}(Q_{Y(0)|G}(\tau|f)|g)$. The rank transformation uses the conditional potential outcome distribution under treatment. I consider only the relative ranks in the interval $[0.01, 0.99]$ and exclude all ranks below $\hat{F}_{Y(0)|G}(0|f)$ (for explanations, see Online Appendix \ref{appA}).}
\end{figure}

\clearpage

\begin{figure}[h]
\caption{Difference between SQTEs of the Job Corps for females and males.}  \label{figa4}
\begin{center}
\includegraphics[width=11cm]{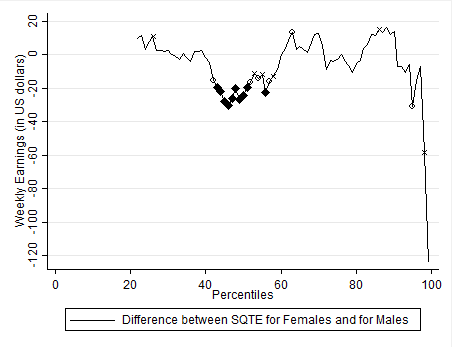}
\end{center}
\parbox{17cm}{\footnotesize Note: The lines report the point estimates calculated separately for all percentiles. Crosses on the lines indicate significant effects at the 10\% level. Hollow circles on the lines indicate significant effects at the 5\% level. Full diamonds on the lines indicate significant effects at the 1\% level. The reference quantiles are from the potential outcome distribution of females under treatment, $Q_{Y(0)|G}(\tau|f)$. The relative rank is defined by $\tau_g^r = F_{Y(1)|G}(Q_{Y(0)|G}(\tau|f)|g)$. The rank transformation uses the conditional potential outcome distribution under treatment. I consider only the relative ranks in the interval $[0.01, 0.99]$ and exclude all ranks below $\hat{F}_{Y(0)|G}(0|f)$ (for explanations, see Online Appendix \ref{appA}).}
\end{figure}

\clearpage

\setcounter{figure}{0}
\setcounter{table}{0}
\section{Distributional heterogeneity by gender and parenthood \label{appE}}

\begin{figure}[h]
\caption{CQTE of the Job Corps program with children by gender on average weekly earnings (in U.S. dollars) in year four after randomized assignment. \label{fi1}} 
\begin{center}
\includegraphics[width=11cm]{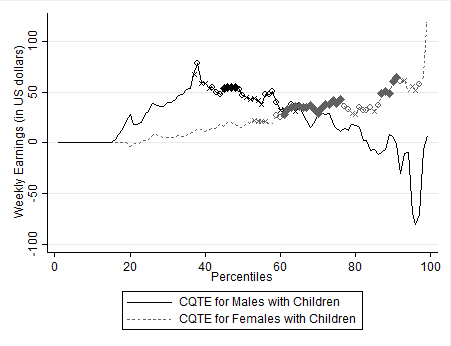}
\end{center}
\parbox{17cm}{\footnotesize Note: The lines report the point estimates calculated separately for all percentiles. Crosses on the lines indicate significant effects at the 10\% level. Hollow circles on the lines indicate significant effects at the 5\% level. Full diamonds on the lines indicate significant effects at the 1\% level.}
\end{figure}

\begin{figure}[h]
\caption{Difference between the Job Corps CQTEs for females and males with children. \label{fi2}} 
\begin{center}
\includegraphics[width=11cm]{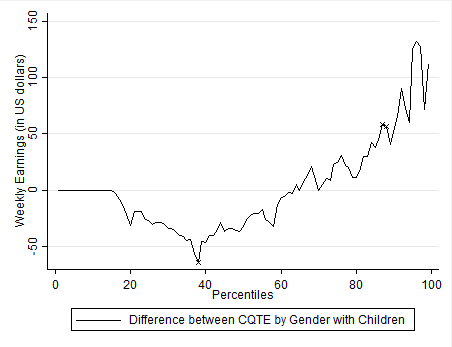}
\end{center}
\parbox{17cm}{\footnotesize Note: The outcome is average weekly earnings (in U.S. dollars) in year four after randomized assignment. The lines report the point estimates calculated separately for all percentiles. Crosses on the lines indicate significant effects at the 10\% level. Hollow circles on the lines indicate significant effects at the 5\% level. Full diamonds on the lines indicate significant effects at the 1\% level.}
\end{figure}

\begin{figure}[h]
\caption{CQTE of the Job Corps program without children by gender on average weekly earnings (in U.S. dollars) in year four after randomized assignment. \label{fi3}} 
\begin{center}
\includegraphics[width=11cm]{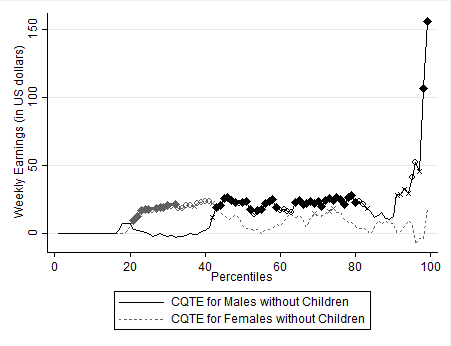}
\end{center}
\parbox{17cm}{\footnotesize Note: The lines report the point estimates calculated separately for all percentiles. Crosses on the lines indicate significant effects at the 10\% level. Hollow circles on the lines indicate significant effects at the 5\% level. Full diamonds on the lines indicate significant effects at the 1\% level.}
\end{figure}

\begin{figure}[h]
\caption{Difference between the Job Corps CQTEs for females and males without children. \label{fi4}} 
\begin{center}
\includegraphics[width=11cm]{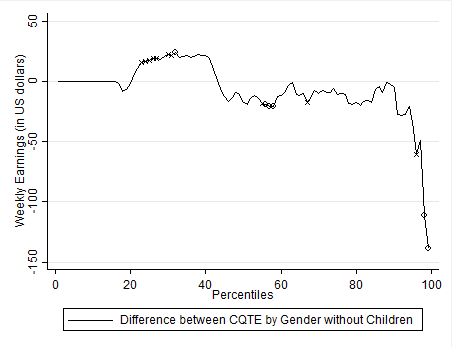}
\end{center}
\parbox{17cm}{\footnotesize Note: The outcome is average weekly earnings (in U.S. dollars) in year four after randomized assignment. The lines report the point estimates calculated separately for all percentiles. Crosses on the lines indicate significant effects at the 10\% level. Hollow circles on the lines indicate significant effects at the 5\% level. Full diamonds on the lines indicate significant effects at the 1\% level.}
\end{figure}

\begin{figure}[h]
\caption{Difference between the Job Corps CQTEs for females and males, with and without children. \label{fi5}} 
\begin{center}
\includegraphics[width=11cm]{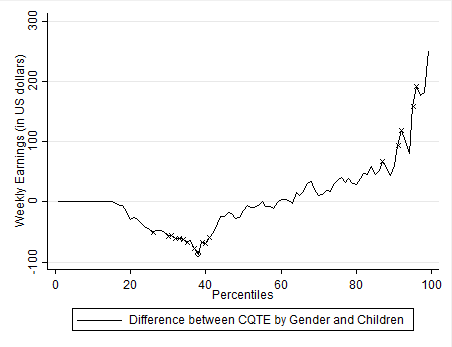}
\end{center}
\parbox{17cm}{\footnotesize Note: The outcome is average weekly earnings (in U.S. dollars) in year four after randomized assignment. The lines report the point estimates calculated separately for all percentiles. Crosses on the lines indicate significant effects at the 10\% level. Hollow circles on the lines indicate significant effects at the 5\% level. Full diamonds on the lines indicate significant effects at the 1\% level.}
\end{figure}

Table \ref{ta1} and Figures \ref{fi1}-\ref{fi5} in Online Appendix \ref{appE} report CQTE by gender and parenthood. Mainly parents and childless men profit from a Job Corps offer. Offering participation in the Job Corps to childless women has the smallest earnings impact of all four groups. The effect heterogeneity by gender is not strongly significant for parents, but significant for childless women and men. Considering heterogeneity between all four groups, the effect heterogeneity by gender and parenthood is large, but the effects are only significant at some quantiles (see Figure \ref{fi5}).

\begin{figure}[h]
\caption{TQTEs of the Job Corps with children by gender on average weekly earnings (in U.S. dollars) in year four after randomized assignment. \label{fi6}} 
\begin{center}
\includegraphics[width=11cm]{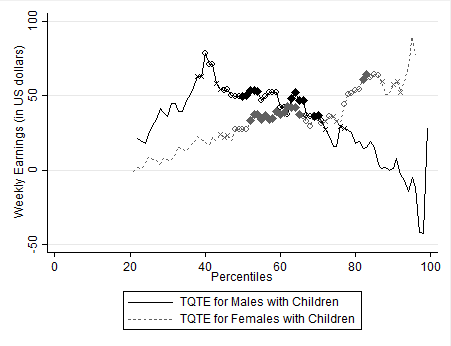}
\end{center}
\parbox{17cm}{\footnotesize Note: The lines report the point estimates separately calculated for all percentiles. Crosses on the lines indicate significant effects at the 10\% level. Hollow circles on the lines indicate significant effects at the 5\% level. Full diamonds on the lines indicate significant effects at the 1\% level. I consider only the relative ranks in the interval $[0.01, 0.99]$ and exclude all ranks below $\hat{F}_{Y(0)}(0)$ (for explanations, see Online Appendix \ref{appA}).}
\end{figure}

\begin{figure}[h]
\caption{Difference between Job Corps TQTEs for females and males with children. \label{fi7}} 
\begin{center}
\includegraphics[width=11cm]{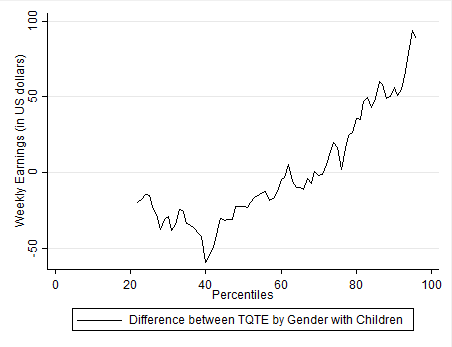}
\end{center}
\parbox{17cm}{\footnotesize Note: The outcome is average weekly earnings (in U.S. dollars) in year four after randomized assignment. The lines report the point estimates separately calculated for all percentiles. All point estimates are not statistically different from zero. I consider the relative ranks only in the interval $[0.01, 0.99]$ and exclude all ranks below $\hat{F}_{Y(0)}(0)$ (for explanations, see Online Appendix \ref{appA}).}
\end{figure}

\begin{figure}[h]
\caption{TQTEs of the Job Corps without children by gender on average weekly earnings (in U.S. dollars) in year four after randomized assignment. \label{fi8}} 
\begin{center}
\includegraphics[width=11cm]{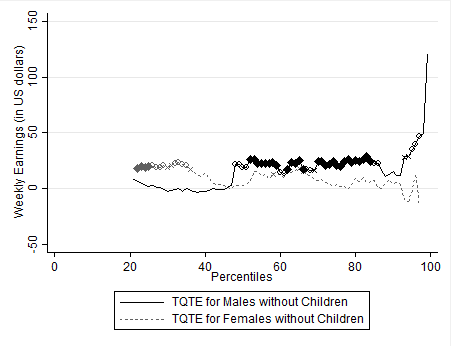}
\end{center}
\parbox{17cm}{\footnotesize Note: The lines report the point estimates separately calculated for all percentiles. Crosses on the lines indicate significant effects at the 10\% level. Hollow circles on the lines indicate significant effects at the 5\% level. Full diamonds on the lines indicate significant effects at the 1\% level. I consider only the relative ranks in the interval $[0.01, 0.99]$ and exclude all ranks below $\hat{F}_{Y(0)}(0)$ (for explanations, see Online Appendix \ref{appA}).}
\end{figure}

\begin{figure}[h]
\caption{Difference between Job Corps TQTEs for females and males without children. \label{fi9}} 
\begin{center}
\includegraphics[width=11cm]{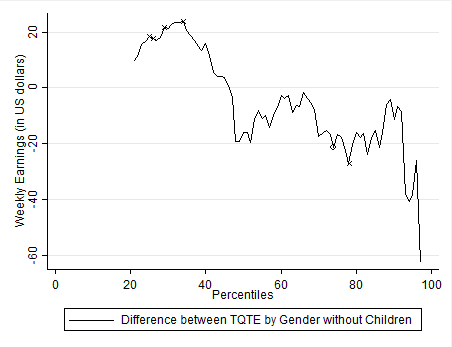}
\end{center}
\parbox{17cm}{\footnotesize Note: The outcome is average weekly earnings (in U.S. dollars) in year four after randomized assignment. The lines report the point estimates separately calculated for all percentiles. All point estimates are not statistically different from zero. I consider the relative ranks only in the interval $[0.01, 0.99]$ and exclude all ranks below $\hat{F}_{Y(0)}(0)$ (for explanations, see Online Appendix \ref{appA}).}
\end{figure}

\begin{figure}[h]
\caption{Difference between Job Corps TQTEs for females and males, with and without children. \label{fi10}} 
\begin{center}
\includegraphics[width=11cm]{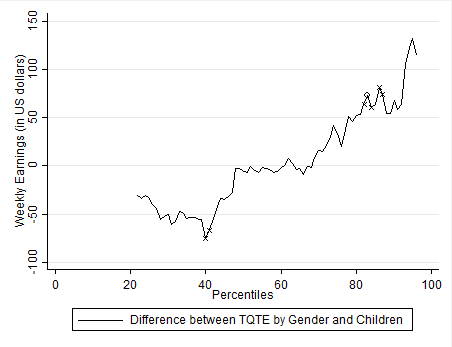}
\end{center}
\parbox{17cm}{\footnotesize Note: The outcome is average weekly earnings (in U.S. dollars) in year four after randomized assignment. The lines report the point estimates separately calculated for all percentiles. All point estimates are not statistically different from zero. I consider the relative ranks only in the interval $[0.01, 0.99]$ and exclude all ranks below $\hat{F}_{Y(0)}(0)$ (for explanations, see Online Appendix \ref{appA}).}
\end{figure}

\begin{figure}[h]
\caption{SQTEs of the Job Corps with children by gender on average weekly earnings (in U.S. dollars) in year four after randomized assignment.\label{fi11}} 
\begin{center}
\includegraphics[width=11cm]{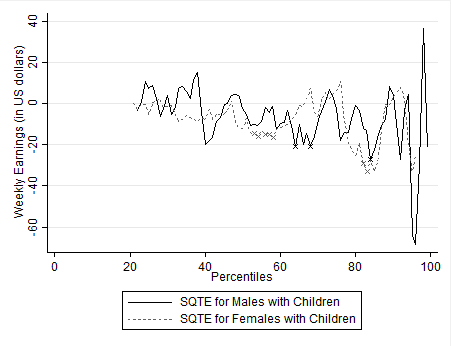}
\end{center}
\parbox{17cm}{\footnotesize Note: The lines report the point estimates separately calculated for all percentiles. Crosses on the lines indicate significant effects at the 10\% level. Hollow circles on the lines indicate significant effects at the 5\% level. Full diamonds on the lines indicate significant effects at the 1\% level. I consider the relative ranks only in the interval $[0.01, 0.99]$ and exclude all ranks below $\hat{F}_{Y(0)}(0)$ (for explanations, see Online Appendix \ref{appA}).}
\end{figure}

\begin{figure}[h]
\caption{Difference between SQTEs of the Job Corps for females and males with children. \label{fi12}} 
\begin{center}
\includegraphics[width=11cm]{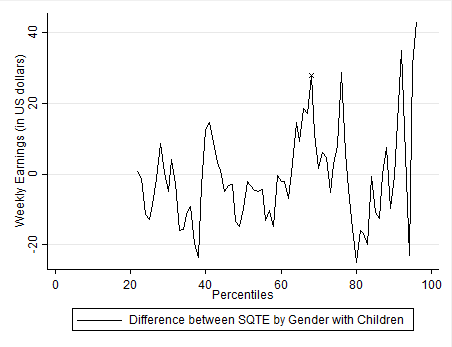}
\end{center}
\parbox{17cm}{\footnotesize Note: The outcome is average weekly earnings (in U.S. dollars) in year four after randomized assignment. The lines report the point estimates separately calculated for all percentiles. Crosses on the lines indicate significant effects at the 10\% level. Hollow circles on the lines indicate significant effects at the 5\% level. Full diamonds on the lines indicate significant effects at the 1\% level. I consider the relative ranks only in the interval $[0.01, 0.99]$ and exclude all ranks below $\hat{F}_{Y(0)}(0)$ (for explanations, see Online Appendix \ref{appA}).}
\end{figure}

\begin{figure}[h]
\caption{SQTEs of the Job Corps without children by gender on average weekly earnings (in U.S. dollars) in year four after randomized assignment. \label{fi13}} 
\begin{center}
\includegraphics[width=11cm]{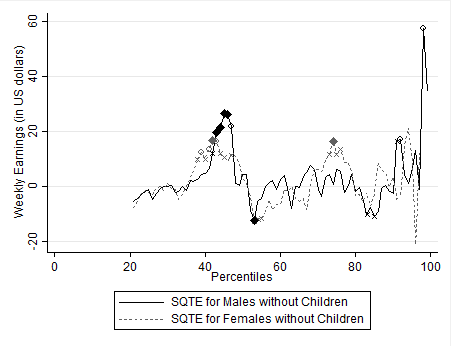}
\end{center}
\parbox{17cm}{\footnotesize Note: The lines report the point estimates separately calculated for all percentiles. Crosses on the lines indicate significant effects at the 10\% level. Hollow circles on the lines indicate significant effects at the 5\% level. Full diamonds on the lines indicate significant effects at the 1\% level. I consider the relative ranks only in the interval $[0.01, 0.99]$ and exclude all ranks below $\hat{F}_{Y(0)}(0)$ (for explanations, see Online Appendix \ref{appA}).}
\end{figure}

\begin{figure}[h]
\caption{Difference between SQTEs of the Job Corps for females and males without children. \label{fi14}} 
\begin{center}
\includegraphics[width=11cm]{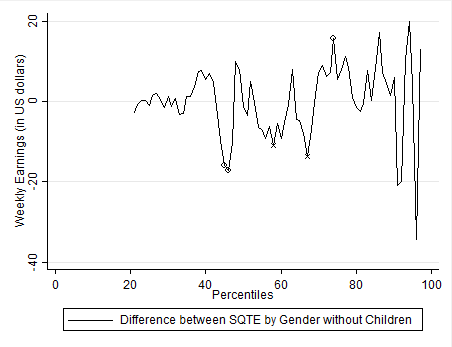}
\end{center}
\parbox{17cm}{\footnotesize Note: The outcome is average weekly earnings (in U.S. dollars) in year four after randomized assignment. The lines report the point estimates separately calculated for all percentiles. Crosses on the lines indicate significant effects at the 10\% level. Hollow circles on the lines indicate significant effects at the 5\% level. Full diamonds on the lines indicate significant effects at the 1\% level. I consider the relative ranks only in the interval $[0.01, 0.99]$ and exclude all ranks below $\hat{F}_{Y(0)}(0)$ (for explanations, see Online Appendix \ref{appA}).}
\end{figure}

\begin{figure}[h]
\caption{Difference between SQTEs of the Job Corps for females and males, with and without children. \label{fi15}} 
\begin{center}
\includegraphics[width=11cm]{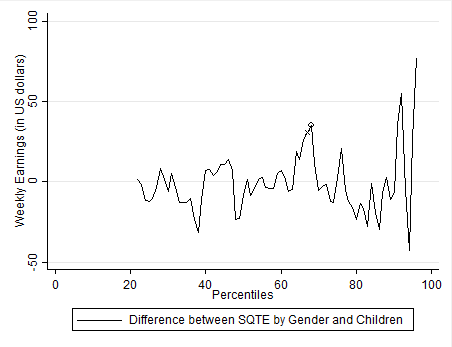}
\end{center}
\parbox{17cm}{\footnotesize Note: The outcome is average weekly earnings (in U.S. dollars) in year four after randomized assignment. The lines report the point estimates separately calculated for all percentiles. Crosses on the lines indicate significant effects at the 10\% level. Hollow circles on the lines indicate significant effects at the 5\% level. Full diamonds on the lines indicate significant effects at the 1\% level. I consider the relative ranks only in the interval $[0.01, 0.99]$ and exclude all ranks below $\hat{F}_{Y(0)}(0)$ (for explanations, see Online Appendix \ref{appA}).}
\end{figure}

Table \ref{ta2} and Figures \ref{fi6}-\ref{fi10} in Online Appendix \ref{appE} report TQTE by gender and parenthood. The TQTE show large effect heterogeneity by gender of parents, but the differences are not significant (see Figure \ref{fi7}). The effect heterogeneity is significant between childless women and men (see Figure \ref{fi9}). Their is a tendency that TQTE can explain heterogeneity between the two groups, because the Kolmogorov--Smirnov test is weakly significant (see Table \ref{ta2} and Figure \ref{fi10}). The results for SQTE are ambiguous (see Table \ref{ta3} and Figures \ref{fi11}-\ref{fi15} in Online Appendix \ref{appE}). We find significant SQTE for all groups, but the signs of the effects are mixed. At some quantile the SQTE have a large magnitude. Accordingly, the TQTE and SQTE contribute both to the effect heterogeneity in the CQTE by gender and parenthood.

\clearpage

\subsection{CQTE by gender and parenthood}

\begin{table}[h]
\caption{CQTE of the Job Corps program by gender and parenthood on average weekly earnings (in U.S. dollars) in year four after randomized assignment.} \label{ta1} \footnotesize
\begin{center}
\begin{tabular}{lccccccc}
\hline
\hline
    Quantile       &   \multicolumn{ 3}{c}{With children} & \multicolumn{ 3}{c}{Without children} & Difference between \\

           &    Females &      Males & Difference &    Females &      Males & Difference & (3) and (6) \\
\cline{2-8}
           &        (1) &        (2) &        (3) &        (4) &        (5) &        (6) &        (7) \\
\hline
0.2 &      -3.41 &       27.8 &     -31.21 &       5.56 &       7.29 &      -1.74 &     -29.48 \\

           &       8.16 &      22.09 &       23.3 &       3.64 &       8.57 &       9.21 &      25.28 \\

0.3 &       5.71 &      39.49 &     -33.78 &    20,5*** &      -2.09 &     22,59* &    -56,38* \\

           &      10.91 &      28.09 &      30.17 &       7.19 &      10.02 &      12.36 &      32.59 \\

0.4 &      11.24 &     58,09* &     -46.85 &    23,95** &       2.49 &      21.46 &    -68,31* \\

           &      11.27 &      30.45 &      32.71 &      10.89 &       9.79 &      14.87 &      36.04 \\

0.5 &      14.97 &      47.02 &     -32.05 &       5.85 &   23,28*** &     -17.43 &     -14.63 \\

           &      14.77 &      22.88 &      27.55 &       11.8 &       7.86 &      14.19 &      31.19 \\

0.6 &    25,49** &       32.4 &      -6.91 &       5.66 &    17,48** &     -11.82 &        4.9 \\

           &       9.97 &      19.44 &      21.95 &       7.11 &       7.67 &       10.6 &       24.7 \\

0.7 &   28,57*** &      28.53 &       0.04 &      13.18 &   23,43*** &     -10.26 &      10.29 \\

           &       9.17 &      18.38 &       20.3 &        9.2 &        4.3 &      10.18 &      22.95 \\

0.8 &     28,01* &      16.88 &      11.13 &       5.81 &   23,21*** &     -17.41 &      28.54 \\

           &      15.66 &      26.84 &      31.08 &      10.19 &       8.75 &      13.39 &      34.09 \\

0.9 &   60,82*** &       5.41 &      55.41 &       7.73 &      13.04 &      -5.31 &      60.73 \\

           &      18.87 &      35.96 &         40 &      15.47 &      12.87 &      20.01 &      45.06 \\
\hline
KS statistic &     12,318 &      8,261 &     13,598 &      2,466 &  16,041*** &    14,295* &     25,861 \\

           &      0.175 &      0.955 &      0.837 &      0.899 &      0.009 &      0.059 &      0.534 \\

PSD statistic &  12,318*** &      7,995 &     13,598 &      2,466 &  16,041*** &      2,513 &     25,861 \\

           &      0.008 &      0.552 &      0.522 &      0.584 &      0.009 &      0.713 &      0.338 \\

NSD statistic &       -351 &     -8,261 &     -6,595 &       -814 &       -265 &  -14,295** &     -8,921 \\

           &       0.97 &      0.621 &      0.674 &      0.856 &      0.966 &      0.025 &      0.587 \\
\hline
\hline
\end{tabular}  
  
\end{center}
\parbox{17cm}{\footnotesize Note: *** indicates significance at the 1\% level, ** indicates significance at the 5\% level, and * indicates significance at the 10\% level. Bootstrapped standard errors are in parentheses and are obtained from 1,999 bootstrap replications. ''KS statistic'' indicates the Kolmogorov--Smirnov statistic. ''PSD statistic'' indicates the Kolmogorov--Smirnov-type test statistic for positive stochastic dominance. ''NSD statistic'' indicates the Kolmogorov--Smirnov-type test statistic for negative stochastic dominance. Bootstrapped p-values (1,999 replications) are in squared brackets.}
\end{table}

\clearpage
\subsection{TQTE by gender and parenthood}

\begin{table}[h]
\caption{TQTEs of the Job Corps by gender and parenthood on average weekly earnings (in U.S. dollars) in year four after randomized assignment.} \label{ta2} \footnotesize
\begin{center}
\begin{tabular}{lccccccc}
\hline
\hline
 Quantile          &   \multicolumn{ 3}{c}{With children} & \multicolumn{ 3}{c}{Without children} & Difference between \\

           &    Females &      Males & Difference &    Females &      Males & Difference & (3) and (6) \\
\cline{2-8}
           &        (1) &        (2) &        (3) &        (4) &        (5) &        (6) &        (7) \\
\hline
0.3 &       6.93 &      35.81 &     -28.88 &      19.29 &      -2.04 &      21.33 &     -50.21 \\

           &      13.54 &      30.88 &      34.02 &     10,74* &       8.22 &      13.67 &      36.69 \\

0.4 &      18.82 &    77,98** &     -59.17 &      14.01 &      -2.07 &      16.08 &     -75.25 \\

           &      15.25 &      32.93 &       36.7 &       9.44 &       9.63 &      13.48 &      39.46 \\

0.5 &    27,16** &     49,6** &     -22.44 &       2.95 &      19.04 &     -16.09 &      -6.35 \\

           &      12.89 &      19.03 &      23.42 &       7.32 &     8,89** &      12.06 &       26.7 \\

0.6 &   37,34*** &    42,13** &      -4.79 &      12.39 &   15,02*** &      -2.63 &      -2.16 \\

           &      11.87 &      20.26 &      23.39 &       10.1 &       5.92 &      11.03 &      25.96 \\

0.7 &    34,97** &    36,52** &      -1.55 &       7.06 &   24,43*** &     -17.37 &      15.82 \\

           &      15.67 &      14.12 &       20.9 &       8.99 &       8.24 &      12.77 &      24.59 \\

0.8 &    53,62** &      17.61 &      36.01 &       9.26 &   25,17*** &     -15.91 &      51.92 \\

           &      21.96 &      18.82 &      28.73 &      12.08 &       8.24 &      14.72 &      32.44 \\

0.9 &     57,41* &       1.35 &      56.06 &       4.54 &      15.76 &     -11.22 &      67.28 \\

           &      30.69 &      29.02 &      42.63 &      21.52 &      14.42 &      24.31 &      49.71 \\
\hline
KS statistic &       9160 &       8027 &       9693 &       2412 &    12488** &       6416 &      13605 \\

           &       0.16 &      0.779 &      0.334 &      0.884 &      0.013 &      0.302 &      0.185 \\

PSD statistic &       9160 &       8027 &       9693 &       2412 &    12488** &       2458 &     13605* \\

           &      0.125 &      0.384 &      0.172 &      0.474 &      0.013 &      0.599 &      0.094 \\

NSD statistic &        -84 &      -4374 &      -6090 &      -1578 &       -303 &      -6416 &      -7745 \\

           &      0.967 &      0.676 &       0.47 &      0.721 &      0.973 &      0.126 &      0.374 \\
\hline
\hline
\end{tabular}  
\end{center}
\parbox{17cm}{\footnotesize Note: *** indicates significance at the 1\% level, ** indicates significance at the 5\% level, and * indicates significance at the 10\% level. Bootstrapped standard errors are in parentheses and are obtained from 1,999 bootstrap replications. ''KS statistic'' indicates the Kolmogorov--Smirnov statistic. ''PSD statistic'' indicates the Kolmogorov--Smirnov-type test statistic for positive stochastic dominance. ''NSD statistic'' indicates the Kolmogorov--Smirnov-type test statistic for negative stochastic dominance. Bootstrapped p-values (1,999 replications) are in squared brackets.}
\end{table}  

\clearpage
\subsection{SQTE by gender and parenthood}

\begin{table}[h]
\caption{SQTEs of the Job Corps by gender and parenthood on average weekly earnings (in U.S. dollars) in year four after randomized assignment.} \label{ta3} \footnotesize
\begin{center}
\begin{tabular}{lccccccc}
\hline
\hline
 Quantile          &   \multicolumn{ 3}{c}{With children} & \multicolumn{ 3}{c}{Without children} & Difference between \\

           &    Females &      Males & Difference &    Females &      Males & Difference & (3) and (6) \\
\cline{2-8}
           &        (1) &        (2) &        (3) &        (4) &        (5) &        (6) &        (7) \\ \hline

0.3 &      -1.22 &       3.68 &       -4.9 &       1.21 &      -0.05 &       1.26 &      -6.16 \\

           &       5.71 &      14.32 &      15.51 &       6.13 &       4.72 &       7.66 &      16.76 \\

0.4 &      -7.57 &      -19.9 &      12.32 &      9,95* &       4.56 &       5.39 &       6.93 \\

           &       8.57 &      16.16 &      18.25 &       6.04 &       5.17 &       7.94 &      19.69 \\

0.5 &      -12.2 &      -2.58 &      -9.62 &        2.9 &       4.24 &      -1.34 &      -8.28 \\

           &       9.48 &      14.08 &      16.38 &       7.59 &       6.86 &       8.77 &      18.65 \\

0.6 &     -11.85 &      -9.73 &      -2.12 &      -6.73 &       2.46 &      -9.18 &       7.06 \\

           &       9.09 &      12.85 &      15.54 &       6.67 &       4.53 &        7.4 &      17.64 \\

0.7 &       -6.4 &      -7.98 &       1.58 &       6.12 &       -1,0 &       7.12 &      -5.53 \\

           &      11.68 &      10.32 &      15.87 &       6.05 &       5.99 &       8.64 &      17.58 \\

0.8 &     -25.61 &      -0.73 &     -24.88 &      -3.45 &      -1.95 &       -1.5 &     -23.38 \\

           &      15.93 &      16.07 &      22.67 &       7.55 &       4.42 &       8.74 &      24.65 \\

0.9 &       3.41 &       4.06 &      -0.65 &       3.19 &      -2.72 &       5.91 &      -6.56 \\

           &      22.83 &       21.3 &      30.49 &      14.85 &       8.32 &      15.94 &      34.73 \\
\hline
KS statistic &       3432 &       7058 &       4446 &       2181 &      5916* &       3522 &       7968 \\

           &       0.89 &      0.918 &      0.973 &      0.905 &      0.096 &      0.738 &      0.679 \\

PSD statistic &       1102 &       3730 &       4446 &       2169 &     5916** &       2047 &       7968 \\

           &      0.993 &       0.88 &      0.742 &      0.726 &      0.046 &       0.88 &      0.369 \\

NSD statistic &      -3432 &      -7058 &      -2561 &      -2181 &      -1263 &      -3522 &      -4404 \\

           &      0.638 &      0.789 &      0.991 &      0.589 &      0.961 &      0.451 &      0.891 \\
\hline
\hline
\end{tabular}  

\end{center}
\parbox{17cm}{\footnotesize Note: *** indicates significance at the 1\% level, ** indicates significance at the 5\% level, and * indicates significance at the 10\% level. Bootstrapped standard errors are in parentheses and are obtained from 1,999 bootstrap replications. ''KS statistic'' indicates the Kolmogorov--Smirnov statistic. ''PSD statistic'' indicates the Kolmogorov--Smirnov-type test statistic for positive stochastic dominance. ''NSD statistic'' indicates the Kolmogorov--Smirnov-type test statistic for negative stochastic dominance. Bootstrapped p-values (1,999 replications) are in squared brackets.}
\end{table}  

\end{appendix}
\end{document}